\documentclass[twocolumn]{aastex62}

\newcommand{\solarmass}{\ensuremath{\mathrm{M}_{\odot}}}

\newcommand{\Msun}{\solarmass}

\newcommand{\MH}{\ensuremath{M_{\mathrm{mol}}}}
\newcommand{\logOH}{\ensuremath{12 + \log{\mathrm{O}/\mathrm{H}}}}

\newcommand{\tdepl}{\ensuremath{t_{\mathrm{depl}}}}

\newcommand{\aco}{\ensuremath{\alpha_{\mathrm{CO}}}}

\newcommand{\Halpha}{\ensuremath{\mathrm{H}\alpha\ \lambda 6563}}

\newcommand{\Oii}{\ensuremath{\mathrm{[O\,\textsc{ii}]}\ \lambda 3726,3729}}
\newcommand{\Oiialt}{\ensuremath{\mathrm{[O\,\textsc{ii}]}\ \lambda 3727}}
\newcommand{\Oiia}{\ensuremath{\mathrm{[O\,\textsc{ii}]}\ \lambda 3726}}

\newcommand{\Oiisum}{\ensuremath{\mathrm{[O\,\textsc{ii}]}\ \lambda 3726 + \lambda3729}}

\newcommand{\Oiii}{\ensuremath{\mathrm{[O\,\textsc{iii}]}\ \lambda 4959,5007}}

\newcommand{\Neiii}{\ensuremath{\mathrm{[Ne\,\textsc{iii}]}\ \lambda 3869}}

\newcommand{\Lyalpha}{\ensuremath{\mathrm{Ly}\alpha\ \lambda 1216}}
\newcommand{\Ciii}{\ensuremath{\mathrm{C\,\textsc{iii}]}\ \lambda 1907, 1909}}

\newcommand{\Mgii}{\ensuremath{\mathrm{Mg\,\textsc{ii}}\ \lambda 2796,2803}}

\newcommand{\Feii}{\ensuremath{\mathrm{Fe\, \textsc{ii}\ \lambda 2586, 2600}}}

\newcommand{\CIII}{\ensuremath{\mathrm{C\,\textsc{iii}]}}}
\newcommand{\MgII}{\ensuremath{\mathrm{Mg\,\textsc{ii}}}}
\newcommand{\FeII}{\ensuremath{\mathrm{Fe\,\textsc{ii}}}}

\newcommand{\Cii}{\ensuremath{\mathrm{C\,\textsc{ii}}\ \lambda 1334}}

\newcommand{\Siiv}{\ensuremath{\mathrm{Si\,\textsc{iv}}\ \lambda 1394, 1402}}

\newcommand{\Siii}{\ensuremath{\mathrm{Si\,\textsc{ii}}\ \lambda 1551}}

\newcommand{\Civ}{\ensuremath{\mathrm{C\,\textsc{iv}}\ \lambda 1548, 1551}}
\newcommand{\Alii}{\ensuremath{\mathrm{Al\,\textsc{ii}}\ \lambda 1671}}
\newcommand{\Aliii}{\ensuremath{\mathrm{Al\,\textsc{iii}}\ \lambda 1855, 1863}}

\newcommand{\Lya}{\ensuremath{\mathrm{Ly}\alpha}}

\newcommand{\Ha}{\ensuremath{\mathrm{H}\alpha}}

\newcommand{\OII}{\ensuremath{\mathrm{[O\,\textsc{ii}]}}}

\newcommand{\NeIII}{\ensuremath{\mathrm{[Ne\,\textsc{iii}]}}}
\newcommand{\CI}{\ensuremath{\mathrm{[C\, \textsc{i}]}}}

\newcommand{\Oisky}{\ensuremath{\mathrm{[O\,\textsc{i}]}\ \lambda 5577}}

\newcommand{\fobs}{\ensuremath{10.8^{+3.0}_{-5.1}}}
\newcommand{\fobsMmedian}{\ensuremath{10^{10.6}}~\Msun}
\newcommand{\fobsprior}{\ensuremath{10.8^{+2.3}_{-5.1}}}
\newcommand{\fobspriorMmedian}{\ensuremath{10^{10.6}}~\Msun}
\newcommand{\fobsall}{\ensuremath{2.2^{+0.2}_{-0.1}}}
\newcommand{\fobsallMmedian}{\ensuremath{10^{9}}~\Msun}

\newcommand{\medTdeplZone}{\ensuremath{\approx 1.2}}
\newcommand{\medTdeplZtwo}{\ensuremath{\approx 1.3}}

\newcommand{\medFgasZone}{\ensuremath{\approx 0.6}}
\newcommand{\medFgasZtwo}{\ensuremath{\approx 2.0}}
\newcommand{\medFgasZthree}{\ensuremath{8.8 \pm 2.8}}

\received{}
\revised{}
\accepted{}

\submitjournal{ApJ}

\shorttitle{ASPECS-LP: Nature \& physical properties of gas-mass selected galaxies}
\shortauthors{Boogaard et al.}

\begin{document}

\title{The ALMA Spectroscopic Survey in the HUDF: Nature and physical
  properties of gas-mass selected galaxies using MUSE spectroscopy}

\correspondingauthor{Leindert Boogaard}
\email{boogaard@strw.leidenuniv.nl}

\author[0000-0002-3952-8588]{Leindert A. Boogaard} \affil{Leiden Observatory,
  Leiden University, PO Box 9513, NL-2300 RA Leiden, The Netherlands}

\author[0000-0002-2662-8803]{Roberto Decarli} \affil{INAF-Osservatorio di
  Astrofisica e Scienza dello Spazio, via Gobetti 93/3, I-40129, Bologna,
  Italy}

\author[0000-0003-3926-1411]{Jorge Gonz\'alez-L\'opez} \affil{N\'ucleo de
  Astronom\'ia de la Facultad de Ingenier\'ia y Ciencias, Universidad Diego
  Portales, Av. Ej\'ercito Libertador 441, Santiago, Chile} \affil{Instituto de
  Astrof\'{\i}sica, Facultad de F\'{\i}sica, Pontificia Universidad Cat\'olica
  de Chile Av. Vicu\~na Mackenna 4860, 782-0436 Macul, Santiago, Chile}

\author{Paul van der Werf} \affil{Leiden Observatory, Leiden University, PO Box
  9513, NL-2300 RA Leiden, The Netherlands}

\author[0000-0003-4793-7880]{Fabian Walter} \affil{Max Planck Institute f\"ur
  Astronomie, K\"onigstuhl 17, 69117 Heidelberg, Germany} \affil{National Radio
  Astronomy Observatory, Pete V. Domenici Array Science Center, P.O. Box O,
  Socorro, NM 87801, USA}

\author{Rychard Bouwens} \affil{Leiden Observatory, Leiden University, PO Box
  9513, NL-2300 RA Leiden, The Netherlands}

\author[0000-0002-6290-3198]{Manuel Aravena} \affil{N\'ucleo de Astronom\'ia de
  la Facultad de Ingenier\'ia y Ciencias, Universidad Diego Portales,
  Av. Ej\'ercito Libertador 441, Santiago, Chile}

\author{Chris Carilli} \affil{National Radio Astronomy Observatory, Pete
  V. Domenici Array Science Center, P.O. Box O, Socorro, NM 87801, USA}
\affil{Battcock Centre for Experimental Astrophysics, Cavendish Laboratory,
  Cambridge CB3 0HE, UK}

\author[0000-0002-8686-8737]{Franz Erik Bauer} \affil{Instituto de
  Astrof\'{\i}sica, Facultad de F\'{\i}sica, Pontificia Universidad Cat\'olica
  de Chile Av. Vicu\~na Mackenna 4860, 782-0436 Macul, Santiago, Chile}
\affil{Millennium Institute of Astrophysics (MAS), Nuncio Monse{\~{n}}or
  S{\'{o}}tero Sanz 100, Providencia, Santiago, Chile} \affil{Space Science
  Institute, 4750 Walnut Street, Suite 205, Boulder, CO 80301, USA}

\author{Jarle Brinchmann} \affil{Leiden Observatory, Leiden University, PO Box
  9513, NL-2300 RA Leiden, The Netherlands} \affil{Instituto de
  Astrof{\'\i}sica e Ci{\^e}ncias do Espa\c{c}o, Universidade do Porto, CAUP, Rua
  das Estrelas, PT4150-762 Porto, Portugal}

\author{Thierry Contini} \affil{Institut de Recherche en Astrophysique et
  Planétologie (IRAP), Université de Toulouse, CNRS, UPS, 31400 Toulouse,
  France}

\author{Pierre Cox} \affil{Institut d'astrophysique de Paris, Sorbonne
  Université, CNRS, UMR 7095, 98 bis bd Arago, 7014 Paris, France}

\author{Elisabete da Cunha} \affil{Research School of Astronomy and
  Astrophysics, Australian National University, Canberra, ACT 2611, Australia}

\author{Emanuele Daddi} \affil{Laboratoire AIM, CEA/DSM-CNRS-Universite Paris
  Diderot, Irfu/Service d'Astrophysique, CEA Saclay, Orme des Merisiers, 91191
  Gif-sur-Yvette cedex, France}

\author{Tanio D\'iaz-Santos} \affil{N\'ucleo de Astronom\'ia de la Facultad de
  Ingenier\'ia y Ciencias, Universidad Diego Portales, Av. Ej\'ercito
  Libertador 441, Santiago, Chile}

\author{Jacqueline Hodge} \affil{Leiden Observatory, Leiden University, PO Box
  9513, NL-2300 RA Leiden, The Netherlands}

\author{Hanae Inami} \affil{Univ. Lyon 1, ENS de Lyon, CNRS, Centre de
  Recherche Astrophysique de Lyon (CRAL) UMR5574, 69230 Saint-Genis-Laval,
  France} \affil{Hiroshima Astrophysical Science Center, Hiroshima University,
  1-3-1 Kagamiyama, Higashi-Hiroshima, Hiroshima, 739-8526}

\author{Rob Ivison} \affil{European Southern Observatory,
  Karl-Schwarzschild-Strasse 2, 85748, Garching, Germany} \affil{Institute for
  Astronomy, University of Edinburgh, Royal Observatory, Blackford Hill,
  Edinburgh EH9 3HJ}

\author{Michael Maseda} \affil{Leiden Observatory, Leiden University, PO Box
  9513, NL-2300 RA Leiden, The Netherlands}

\author{Jorryt Matthee} \affil{Department of Physics, ETH Zurich,
  Wolfgang-Pauli-Strasse 27, 8093, Zurich, Switzerland}

\author{Pascal Oesch} \affil{Department of Astronomy, University of Geneva,
  Ch. des Maillettes 51, 1290 Versoix, Switzerland}

\author{Gerg\"{o} Popping} \affil{Max Planck Institute f\"ur Astronomie,
  K\"onigstuhl 17, 69117 Heidelberg, Germany}

\author[0000-0001-9585-1462]{Dominik Riechers} \affil{Cornell University, 220
  Space Sciences Building, Ithaca, NY 14853, USA} \affil{Max Planck Institute
  f\"ur Astronomie, K\"onigstuhl 17, 69117 Heidelberg, Germany}

\author{Joop Schaye} \affil{Leiden Observatory, Leiden University, PO Box
  9513, NL-2300 RA Leiden, The Netherlands}

\author{Sander Schouws} \affil{Leiden Observatory, Leiden University, PO Box
  9513, NL-2300 RA Leiden, The Netherlands}

\author{Ian Smail} \affil{Centre for Extragalactic Astronomy, Department of
  Physics, Durham University, South Road, Durham, DH1 3LE, UK}

\author{Axel Weiss} \affil{Max-Planck-Institut f\"ur Radioastronomie, Auf dem
  H\"ugel 69, 53121 Bonn, Germany}

\author{Lutz Wisotzki} \affil{Leibniz-Institut für Astrophysik Potsdam, An der
  Sternwarte 16, 14482 Potsdam, Germany}

\author{Roland Bacon} \affil{Univ. Lyon 1, ENS de Lyon, CNRS, Centre de
  Recherche Astrophysique de Lyon (CRAL) UMR5574, 69230 Saint-Genis-Laval,
  France}

\author{Paulo C.~Cortes} \affil{Joint ALMA Observatory - ESO, Av. Alonso de
  C\'ordova, 3104, Santiago, Chile} \affil{National Radio Astronomy
  Observatory, 520 Edgemont Rd, Charlottesville, VA, 22903, USA}

\author{Hans--Walter Rix} \affil{Max Planck Institute f\"ur Astronomie,
  K\"onigstuhl 17, 69117 Heidelberg, Germany}

\author{Rachel S. Somerville} \affil{Department of Physics and Astronomy,
  Rutgers, The State University of New Jersey, 136 Frelinghuysen Rd,
  Piscataway, NJ 08854, USA} \affil{Center for Computational Astrophysics,
  Flatiron Institute, 162 5th Ave, New York, NY 10010, USA}

\author{Mark Swinbank} \affil{Centre for Extragalactic Astronomy, Department of
  Physics, Durham University, South Road, Durham, DH1 3LE, UK}

\author{Jeff Wagg} \affil{SKA Organization, Lower Withington Macclesfield,
  Cheshire SK11 9DL, UK}

\begin{abstract} %
  We discuss the nature and physical properties of gas-mass selected galaxies
  in the ALMA spectroscopic survey (ASPECS) of the Hubble Ultra Deep Field
  (HUDF).  We capitalize on the deep optical integral-field spectroscopy from
  the MUSE HUDF Survey and multi-wavelength data to uniquely associate all 16
  line-emitters, detected in the ALMA data without preselection, with
  rotational transitions of carbon monoxide (CO).  We identify ten as CO(2-1)
  at $1 < z < 2$, five as CO(3-2) at $2 < z < 3$ and one as CO(4-3) at $z=3.6$.
  Using the MUSE data as a prior, we identify two additional CO(2-1)-emitters,
  increasing the total sample size to 18.  We infer metallicities consistent
  with (super-)solar for the CO-detected galaxies at $z\le1.5$, motivating our
  choice of a Galactic conversion factor between CO luminosity and molecular
  gas mass for these galaxies.  Using deep \emph{Chandra} imaging of the HUDF,
  we determine an X-ray AGN fraction of 20\% and 60\% among the CO-emitters at
  $z\sim1.4$ and $z\sim2.6$, respectively.  Being a CO-flux limited survey,
  ASPECS-LP detects molecular gas in galaxies on, above and below the main
  sequence (MS) at $z\sim1.4$.  For stellar masses
  $\ge 10^{10} (10^{10.5})$~\Msun, we detect about 40\% (50\%) of all galaxies
  in the HUDF at $1 < z < 2$ ($2 < z < 3$).  The combination of ALMA and MUSE
  integral-field spectroscopy thus enables an unprecedented view on MS galaxies
  during the peak of galaxy formation.
\end{abstract}

\keywords{galaxies: high-redshift --- galaxies: ISM --- galaxies: star
  formation}

\section{Introduction}
\label{sec:intro}
Star formation takes place in the cold interstellar medium (ISM) and studying
the cold molecular gas content of galaxies is therefore fundamental for our
understanding of the formation and evolution of galaxies.  As there is little
to no emission from the molecular hydrogen that constitutes the majority of the
molecular gas in mass, cold molecular gas is typically traced by molecules,
such as the bright rotational transitions of $^{12}$C$^{16}$O (hereafter CO).

Recent years have seen a tremendous advance in the characterization of the
molecular gas content of high redshift galaxies \citep[for a review,
see][]{Carilli2013}.  Targeted surveys with the Atacama Large Millimetre Array
(ALMA) and the Plateau de Bure Interferometer (PdBI) have been instrumental in
our understanding of the increasing molecular gas reservoirs of star-forming
galaxies at $z>1$ \citep{Daddi2010, Daddi2015, Genzel2010, Tacconi2010,
  Tacconi2013, Silverman2015, Silverman2018}.  Combining data across cosmic
time, these provide constraints on how the molecular gas content of galaxies
evolves as a function of their physical properties, such as stellar mass
($M_{*}$) and star formation rate (SFR) \citep{Scoville2014, Scoville2017,
  Genzel2015, Saintonge2016, Tacconi2013, Tacconi2018}.  These surveys
typically target galaxies with SFRs that are greater than or equal to the
majority of the galaxy population at their respective redshifts and stellar
masses (the `main sequence' of star-forming galaxies; \citealt{Brinchmann2004,
  Noeske2007a, Whitaker2014, Schreiber2015, Eales2018, Boogaard2018}), and
therefore should be complemented by studies that do not rely on such a
preselection.

Spectral line scans in the (sub-)millimeter regime in deep fields provide a
unique window on the molecular gas content of the universe.  As the cosmic
volume probed is well defined, they play a fundamental role in determining the
evolution of the cosmic molecular gas density through cosmic time.  Through
their spectral scan strategy, these surveys are designed to detect molecular
gas in galaxies without any preselection, providing a flux-limited view on the
molecular gas emission at different redshifts \citep{Walter2014, Decarli2014,
  Walter2016, Decarli2016, Riechers2018, Pavesi2018}.  By conducting
`spectroscopy-of-everything', these can in principle reveal the molecular gas
content in galaxies that would not be selected in traditional studies (e.g.,
galaxies with a low SFR, well below the main sequence, but a substantial gas
mass.).

This paper is part of series of papers presenting the first results from the
ALMA Spectroscopic Survey Large Program (ASPECS-LP; \citealt{DecarliALP}).  The
ASPECS-LP is a spectral line scan targeting the \emph{Hubble} Ultra Deep Field
(HUDF).  Here we use the results from the spectral scan of Band 3 (84-115 GHz;
3.6-2.6 mm) and investigate the nature and physical properties of galaxies
detected in molecular emission lines by ALMA.  In order to do so, it is
important to know about the physical conditions of the galaxies detected in
molecular gas, such as their ISM conditions, their (\emph{HST}) morphology and
stellar and ionized gas dynamics.  The HUDF benefits from the deepest and most
extensive multi-wavelength data, and as of recently, ultra-deep integral-field
spectroscopy.

A critical step in identifying ALMA emission lines with actual galaxies relies
on matching the galaxies in redshift.  In this context, the Multi Unit
Spectroscopic Explorer (MUSE, \citealt{Bacon2010}) HUDF survey, that provides a
deep optical integral-field spectroscopic survey over the HUDF
\citep{Bacon2017}, is essential.  The MUSE HUDF is a natural complement to the
ASPECS-LP in the same area on the sky, providing optical spectroscopy for all
galaxies within the field of view, also without any preselection.  In addition,
the integral-field spectrograph provides redshifts for over a thousands of
galaxies in the HUDF (increasing the number of previously known redshifts by a
factor $\sim10\times$; \citealt{Inami2017}).  Depending on the redshift, these
data can provide key information on the ISM conditions (such as metallicity and
dynamics) of the galaxies harboring molecular gas.  As we will see throughout
this paper, the MUSE data are a significant step forward in our understanding
of galaxy population selected with ALMA.

The paper is organized as follows: We first introduce the spectroscopic and
multi-wavelength data (\autoref{sec:observations}).  We discuss the redshift
identification of the CO-detected galaxies from the line search
\citep{Gonzalez-LopezALP}, using the MUSE and multi-wavelength data, in
\autoref{sec:redsh-ident}.  Next, we leverage the large number of MUSE
redshifts to separate real from spurious sources down to a significantly lower
signal-to-noise ratios (S/N) than possible in the line search
(\autoref{sec:addit-aspecs-sourc}).  Together, these sources form the full
ASPECS-LP Band 3 sample (\autoref{sec:combined-sample}).  We then move on to
the central question(s) of this paper: By doing a survey of molecular gas, in
what kind of galaxies do we detect molecular gas emission at different
redshifts, and what are the physical properties of these galaxies?  We
determine stellar masses, SFRs and (where possible) metallicities for all
sources in (\autoref{sec:deriv-phys-prop}) and link these to the molecular gas
content (\MH) to derive the gas fraction ($\MH/M_{*}$, the molecular-to-stellar
mass ratio) and depletion time ($\tdepl = \MH / \mathrm{SFR}$).  We first
discuss the properties of the sample of CO-detected galaxies in the context of
the overall population of the HUDF (\autoref{sec:stellar-mass-sfr}) and
investigate the X-ray AGN fraction among the detected sources
(\autoref{sec:agn-fraction}).  Using the MUSE spectra, we determine the
unobscured SFR (\autoref{sec:star-formation-rates}) and the metallicity of the
$1 < z < 1.5$ sources (\autoref{sec:metallicities-at-1.0}).  Finally, we
discuss the CO detected galaxies from the flux-limited survey in the context of
the galaxy main sequence (\autoref{sec:discussion}), focusing on the molecular
gas mass, gas fraction and depletion time.  We discuss what fraction of the
galaxy population in the HUDF we detect with increasing redshift.  A further
discussion of the molecular gas properties of these sources data will be
presented in \cite{AravenaALP}.

Throughout this paper, we adopt a \cite{Chabrier2003} IMF and a flat
$\Lambda$CDM cosmology, with $H_{0} = 70$ km~s$^{-1}$ Mpc$^{-1}$,
$\Omega_{m} = 0.3$ and $\Omega_{\Lambda} = 0.7$.  Magnitudes are in the AB
system \citep{Oke1983}.

\section{Observations} \label{sec:observations}

\subsection{ALMA Spectroscopic Survey}
\label{sec:alma}
We focus on the ASPECS-LP Band 3 observations, that have been completed in ALMA
Cycle 4.  The acquisition and reduction of the Band 3 data are described in
detail in \cite{DecarliALP}.  The final mosaic covers a 4.6~arcmin$^{2}$ area
in the HUDF (where the primary beam response is $>50\%$ of the peak
sensitivity).  The data are combined into a single spectral cube with a spatial
resolution of $\approx 1.75\arcsec \times 1.49\arcsec$ (synthesized beam with
natural weighting at 99.5~GHz) and a spectral resolution of 7.813 MHz,
corresponding to $\Delta v \approx 23.5$ km~s$^{-1}$ at 99.5 GHz.  The average
root mean square (rms) sensitivity is $\approx 0.2$ mJy beam$^{-1}$ but varies
across the frequency range, being deepest ($\approx 0.13$ mJy beam$^{-1}$)
around 100 GHz and higher above 110 GHz, due to the spectral setup of the
observations (see \citealt{Gonzalez-LopezALP} for details).  Throughout this
paper, we consider the area that lies within $>40\%$ of the primary beam peak
sensitivity, which is the shallowest part of the survey over which we still
detect CO candidates without preselection (\autoref{sec:redsh-ident}).  When
comparing to the \emph{HST} reference frame, we take into account an
astrometric offset of
$\Delta \alpha = +0.076\arcsec , \Delta \delta = - 0.279\arcsec$
\citep{Dunlop2017,Rujopakarn2016}.

We perform an extensive search of the cube for molecular emission lines, as is
detailed in \cite{Gonzalez-LopezALP} and \autoref{sec:methods}.  With the Band
3 data alone, the ASPECS-LP is sensitive to different CO and \CI\ transitions
at specific redshift ranges which are indicated in the top panel of
\autoref{fig:MUSE-zdist}.

\subsection{MUSE HUDF Survey}
\label{sec:muse}
\begin{figure}[t]
  \includegraphics[width=\columnwidth]{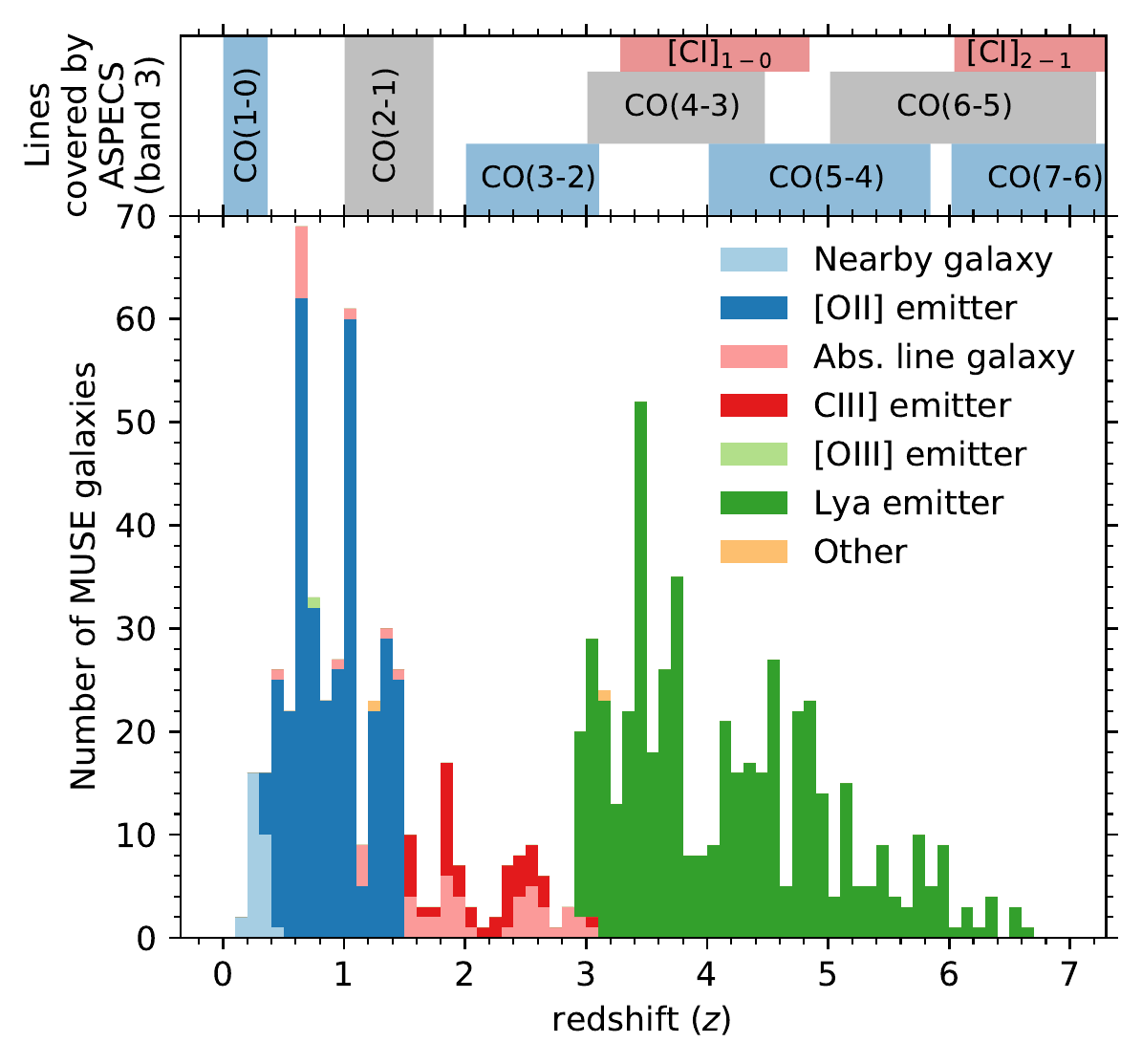}
  \caption{The molecular line redshift coverage of the galaxies in the MUSE and
    ASPECS-LP \emph{Hubble} Ultra Deep Field (HUDF).  The histogram shows the
    galaxies with spectroscopic redshifts from MUSE (\emph{udf10} and
    \emph{mosaic}; see \autoref{sec:muse}) that lie within $>40\%$ of the
    primary beam sensitivity of the ASPECS-LP mosaic, distinguished by the
    primary spectral feature used to identify the redshift
    (\citealt{Inami2017}; `Nearby galaxy' summarizes a range of rest-frame
    optical features).  The decrease in the number of redshifts between
    $1.5 < z < 2.9$ is due to the lack of strong emission line features in the
    MUSE spectrograph (`redshift desert').  The drop at the lowest redshifts is
    due to the nature and volume of the HUDF.  The top panel shows the specific
    CO and \CI\ transitions covered by the frequency setup of ASPECS Band 3 at
    different redshifts \citep{Walter2016, DecarliALP}.  ASPECS covers CO(2-1)
    for \OII\ emitters and absorption line galaxies at $1.0 < z < 1.74$.
    Galaxies with CO(3-2) at $2.0 < z < 3.11$ are identified mostly by UV
    absorption and weaker emission lines (e.g., \CIII).  For higher-order CO
    and \CI\ transitions above $z>2.90$, MUSE has coverage of
    \Lya.\label{fig:MUSE-zdist}}
\end{figure}

The HUDF was observed with the MUSE as part of the MUSE \emph{Hubble} Ultra
Deep Field survey \citep{Bacon2017}.  The location on the sky of the ASPECS-LP
with respect to the MUSE HUDF is shown in \cite{DecarliALP}, Fig. 1.  The MUSE
integral-field spectrograph has a $1' \times 1'$ field-of-view, covering the
optical regime ($4750 - 9300$\AA) at an average spectral resolution of
$\lambda/\Delta\lambda \approx 3000$.  The HUDF was observed in a two tier
strategy, with the \emph{mosaic}-region reaching a median depth of 10 hours in
a $3' \times 3'$-region and the \emph{udf10}-pointing reaching 31 hours depth
in a $1' \times 1'$-region (3$\sigma$ emission line depth for a point source of
3.1 and 1.5~$\times 10^{-19}$~erg~s$^{-1}$~cm$^{-2}$ at 7000\AA, respectively).
The data acquisition and reduction as well as the automated source detection
are described in detail in \cite{Bacon2017}.  The measured seeing in the
reduced datacube is $0\farcs65$ full-width at half-maximum (FWHM) at 7000\AA.

Redshifts were identified semi-automatically and the full spectroscopic catalog
is presented in \cite{Inami2017}.  The spectra were extracted using a weighted
extraction, where the weighting was based on the MUSE white light image, to
obtain the maximal signal-to-noise.  The spectra are modeled with a modified
version of \textsc{platefit} \citep{Tremonti2004, Brinchmann2004,
  Brinchmann2008} to obtain line-flux measurements and equivalent widths for
all sources.  The typical uncertainty on the redshift measurement is
$\sigma_{v} = 0.00012(1+z)$ or $\approx 40$ km~s$^{-1}$ \citep{Inami2017},
which we use to compute the uncertainties in the relative velocities.

In order to compare in detail the relative velocities measured between the
UV/optical features in MUSE and CO in ALMA, we need to place both on the same
reference frame.  The MUSE redshifts are provided in the barycentric reference
frame, while the ALMA cube is set to the kinematic local standard of rest
(LSRK).  When determining detailed velocity offsets we place both on the same
reference frame by removing the velocity difference;
$\mathrm{BARY} - \mathrm{LSRK} = -16.7$~km~s$^{-1}$ (accounting for the angle
between the LSRK vector and the observation direction towards the HUDF).

The redshift distribution of the MUSE galaxies that fall within $>40\%$ of the
primary beam peak sensitivity of the ASPECS-LP footprint in the HUDF is shown
in \autoref{fig:MUSE-zdist}, where galaxies are color coded by the primary
spectral feature(s) used to identify the redshift (see \citealt{Inami2017} for
details).  The redshifts that correspond to the ASPECS band 3 coverage of the
different molecular lines are indicated in the top panel.  CO(1-0)~[115.27~GHz]
is observable at the lowest redshifts ($z < 0.3694$), where MUSE still covers a
major part of the rest-frame optical spectrum that contains a wealth of
spectral features, including absorption and (strong) emission lines (e.g.,
\Halpha, \Oiii\ and \Oii).  The strong lines are the main spectral features
used to identify star-forming galaxies all the way up to $z<1.50$, where \Oii\
moves out of the spectral range of MUSE.  CO(2-1)~[230.54~GHz] is covered by
ASPECS at $1.0059 < z < 1.7387$, mostly overlapping with \OII\ in MUSE.  At
$z>1.5$, the main features used to identify these galaxies are absorption lines
such \Mgii\ and \Feii.  Over the redshift range of CO(3-2)~[345.80~GHz],
$2.0088 < z < 3.1080$, MUSE only has coverage of weaker UV emission lines
(mainly \Ciii), making redshift identifications more challenging (the `redshift
desert').  Here, UV absorption lines are commonly used to identify redshifts,
for galaxies where the continuum is strong enough
($m_{\mathrm{F775W}} \lesssim 26$ mag).  Above $z = 2.9$, MUSE flourishes
again, with the coverage of \Lyalpha\ all the way out to $z\approx 6.7$.  Here,
ASPECS covers CO(4-3)~[461.04~GHz] and transitions with $J_{\rm up} \ge 4$, and
atomic carbon lines ($\CI_{1-0}$~610~\micron\ and $\CI_{2-1}$~370~\micron).

\subsection{Multi-wavelength data (UV--radio) and \textsc{Magphys}}
\label{sec:multi-wavel-data}
In order to construct spectral energy distributions (SEDs) for the ASPECS-LP
sources, we utilize the wealth of available photometric data over the HUDF,
summarized below.

We use the photometric compilation by \citet[][see references
therein]{Skelton2014}, which includes UV, optical and near-IR photometry from
the \emph{Hubble} Space Telescope (\emph{HST}) and ground-based facilities, as
well as (deblended) \emph{Spitzer}/IRAC 3.6\micron, 4.5\micron, 5.8\micron\ and
8.0\micron.  We also include the corresponding deblended \emph{Spitzer}/MIPS
24\micron\ photometry from \citep{Whitaker2014}.  We take deblended
far-infrared (FIR) data from \emph{Herschel}/PACS 100\micron\ and 160\micron\
from \cite{Elbaz2011}, which have a native resolution of 6\farcs7 and
11\farcs0, respectively.  The PACS 100\micron\ and 160\micron\ have a $3\sigma$
depth of 0.8 mJy and 2.4 mJy and are limited by confusion.  For the flux
uncertainties we use the maximum of the local and simulated noise levels for
each source, as recommended by the
documentation\footnote{\url{https://hedam.lam.fr/GOODS-Herschel/data/files/documentation/GOODS-Herschel_release.pdf}}.
We further include the 1.2~mm continuum data from the combination of the
available ASPECS-LP data with the ALMA observations by \cite{Dunlop2017}, taken
over the same region, as detailed in \cite{AravenaALP}.  We also include the
ASPECS-LP 3.0~mm continuum data, as presented in \citep{Gonzalez-LopezALP}.
For the ASPECS survey we have created a master photometry catalog for the
galaxies in the HUDF, adopting the spectroscopic redshifts from MUSE
(\autoref{sec:muse}) and literature sources, as detailed in \cite{DecarliALP}.

We use the high-$z$ extension of the SED-fitting code \textsc{magphys} to infer
physical parameters from the photometric information of the galaxies in our
field \citep{DaCunha2008,DaCunha2015}.  The high-$z$ extension of
\textsc{magphys} includes a larger library of spectral emission models that
extend to higher dust optical depths, higher SFRs and younger ages compared to
what is typically found in the local universe.  From the spectral emission
models, the code can constrain the stellar mass, sSFR and the dust attenuation
($A_{V}$) along the line of sight.  An energy balance argument ensures that the
amount of absorption at rest-frame UV/optical wavelengths is consistent with
the light reradiated in the infrared.  The code performs a Bayesian inference
of the posterior likelihood distribution of the fitted parameter, to account
for uncertainties such as degeneracies in the models, missing data and
non-detections.

We run \textsc{magphys} on all the galaxies in our catalog, using the available
photometric information in all the bands (listed in
\autoref{sec:magphys-fits}).  We do not include the {\em Spitzer}/MIPS and {\em
  Herschel}/PACS photometry in the fits of the general sample because the
angular resolution of these observations is relatively modest ($>5\arcsec$),
thus a delicate de-blending analysis would be required (the average sky density
of galaxies in the HUDF is $\gtrsim$ 1 galaxy per 3 arcsec$^2$).  For the
CO-detected galaxies we repeat the \textsc{magphys} fits including these bands
(\autoref{sec:sfrs}).  In order to take into account systematic errors in the
zero point fitting for these sources, we add the zero point errors
\citep{Skelton2014} in quadrature to the flux errors in all filters except
\emph{HST}, and include a 5\% error-floor to further account for systematic
errors in the physical models (following \citealt{Leja2018}).  The filter
selection of the general sample provides excellent photometric coverage of the
stellar population.  Paired with the wealth of spectroscopic redshifts (see
\citealt{DecarliALP} for a detailed description), this enables robust
constraints on properties such as $M_*$, SFR and $A_{V}$.  We do note that
while the formal uncertainties on the inferred properties are generally small,
systematic uncertainties can be of order $\sim 0.3$~dex
\citep[e.g.,][]{Conroy2013}.

\subsection{X-ray photometry}
\label{sec:x-ray-photometry}
To identify AGN in the field, we use the \emph{Chandra} X-ray data available
over the GOODS-S region from \cite{Luo2017}, which reaches the full depth of 7
Ms over the HUDF area.  In total, there are 36 X-ray sources within the
ASPECS-LP region of the HUDF (i.e., within 40\% of the primary beam).  We
spatially cross-match the X-ray catalog to the closest source within 1\arcsec\
in our MUSE and multi-wavelength catalog over the ASPECS-LP area, visually
inspecting all matches used in this paper to ensure they are accurately
identified.

At the depth of the X-ray data, there are multiple physical mechanisms (e.g.,
AGN and star formation) that may produce the X-ray emission detected at
$0.5 - 7$ keV.  \cite{Luo2017} adopt the following 6 criteria to distinguish
X-ray AGN from other sources of X-ray emission, of which at least one needs to
be satisfied to be classified as AGN (we refer the reader to \citealt{Xue2011},
\citealt{Luo2017} and references therein for details): (1)
$L_{X} \geq 3 \times 10^{42}$ erg s$^{-1}$, identifying luminous X-ray sources;
(2) an effective photon index $\Gamma_{\mathrm{eff}} \leq 1.0$ indicating hard
X-ray sources, identifying obscured AGN; (3) X-ray-to-R-band flux ratio of
$\log\left(f_{X}/f_{R}\right) > -1$; (4) spectroscopically classified as AGN
via, e.g., broad emission lines and/or high excitation lines; (5)
X-ray-to-radio flux ratio of
$L_{X}/L_{1.4 \mathrm{GHz}} \geq 2.4 \times 10^{18}$, indicating an excess of
X-ray emission over the level expected from pure star formation; (6)
X-ray-to-K-band flux ratio of $\log\left(f_{X} / f_{K_{s}}\right) > -1.2$.
Note that even with these criteria it is possible that some X-ray sources host
low-luminosity or heavily obscured AGN and are currently misclassified.

Overall, there are six X-ray AGN in the ASPECS-LP volume at $1.0 < z < 1.7$,
all of which have a MUSE redshift (one being a broad-line AGN).  In the
ASPECS-LP volume at $2.0 < z < 3.1$, there are seven X-ray AGN, three of which
have spectroscopic redshifts from MUSE (including one broad-line AGN), and four
with a photometric redshift (we discard one source in the catalog with a
photometric redshift in this regime for which we cannot securely identify a
counterpart in \emph{HST}).  There is one X-ray AGN at a higher redshift, which
is also identified by MUSE as a broad-line AGN at $z=3.188$.

\section{The ASPECS-LP sample}
\label{sec:methods}

\subsection{Identification of the line search sample}
\label{sec:redsh-ident}

\begin{figure*}[t]
  \includegraphics[width=\textwidth]{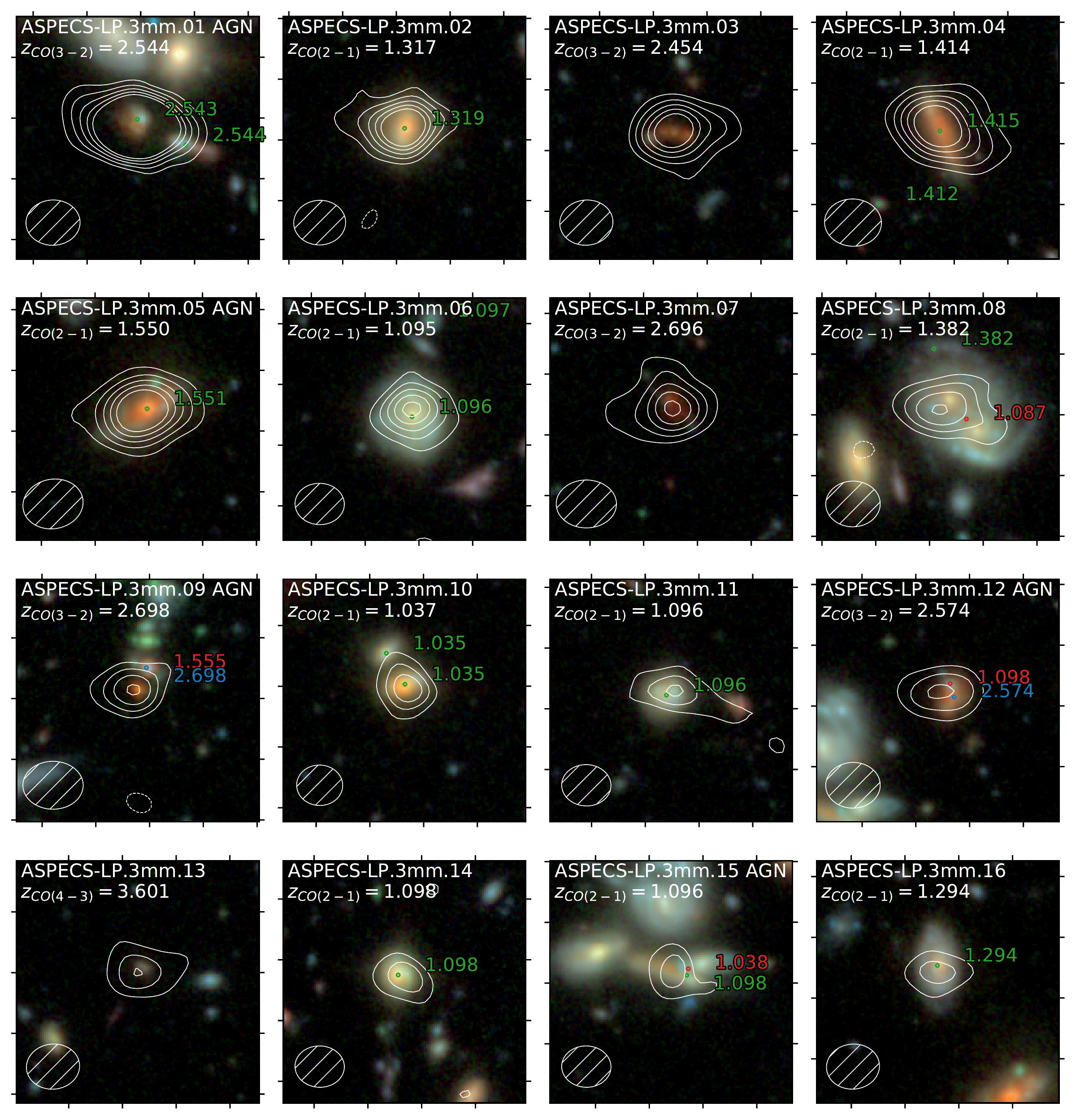}
  \caption{\emph{HST} RGB cutouts (F160W, F125W, F105W) of the 16 CO line
    detections from the line search, all revealing an optical/NIR counterpart.
    Each panel is $8 \times 8 $ arcsec centered around the CO emission
    (corrected astrometry; \autoref{sec:alma}).  The white contours indicate
    the CO signal from $\pm[3, .., 11]\sigma$ in steps of $2\sigma$.  The ALMA
    beam indicated in the bottom left corner.  Galaxies with a spectroscopic
    redshift from MUSE \citep{Inami2017} matching the CO signal are labeled in
    green (and red if not matching); spectroscopic redshifts in blue are newly
    determined in this paper.  Of the 16 galaxies, 12 match closely to a
    redshift from MUSE (including ASPECS-LP.3mm.08, discussed in
    \autoref{sec:redsh-conf} and \citealt{Decarli2016}).  ASPECS-LP.3mm.03,
    3mm.07 and 3mm.09 have $m_{\mathrm{F775W}} > 27$, which is too faint for a
    direct absorption line redshift from MUSE (but are independently
    confirmed).  For ASPECS-LP.3mm.09 we do find UV absorption features
    matching the CO(3-2) in the galaxy slightly to the north.  A new absorption
    line redshift is found for ASPECS-LP.3mm.12 (see \autoref{fig:ASPECS12}).
    The photometric redshift and absence of lower-$z$ spectral features
    indicate ASPECS-LP.3mm.13 being at
    $z=3.601$. \label{fig:HST-cutouts-CO-det}}
\end{figure*}

\begin{table*}[t]
  \centering
  \caption{ASPECS-LP CO detected sources from the line search, with MUSE
    spectroscopic counterparts.  The CO frequencies are taken from
    \citet[][their Table 7]{Gonzalez-LopezALP}. (1) ASPECS-LP 3mm ID. (2)-(3)
    Coordinates. (4) CO line Frequency.  (5) Identified CO transition
    (\autoref{sec:redsh-ident}). (6) CO redshift.  (7) MUSE ID. (8) MUSE
    redshift. (9) Velocity offset between MUSE and ALMA
    ($\Delta v = (z_{\rm MUSE} - z_{\rm CO})/(1 + z_{\rm CO})$; after
    converting both to the same reference frame). \label{tab:CO-line-iden}}
  \begin{tabular}{ccccccccc}
    \hline\hline
    ID & R.A.    & Dec.   & $\nu_{\mathrm{CO}}$ & CO trans. & $z_{\mathrm{CO}}$ & MUSE ID & $z_{\mathrm{MUSE}}$ & $\Delta v$\\
    & (J2000) & (J2000) & (GHz) & ($J_{\rm up}\rightarrow J_{\rm low}$) & & & & km~s$^{-1}$\\
    (1) & (2) & (3) & (4) & (5) & (6) & (7) & (8) & (9) \\
    \hline
    3mm.01 & 03:32:38.54 & -27:46:34.6 & $97.584 \pm 0.003$ & $3\rightarrow2$ & 2.5436 & 35 & 2.5432 & $-15.5 \pm 41.0$ \\
    3mm.02 & 03:32:42.38 & -27:47:07.9 & $99.510 \pm 0.005$ & $2\rightarrow1$ & 1.3167 & 996 & 1.3172$^{*}$ & $73.5 \pm 42.7$ \\
    3mm.03 & 03:32:41.02 & -27:46:31.5 & $100.131 \pm 0.005$ & $3\rightarrow2$ & 2.4534 & \nodata & \nodata & \nodata \\
    3mm.04 & 03:32:34.44 & -27:46:59.8 & $95.501 \pm 0.006$ & $2\rightarrow1$ & 1.4140 & 1117 & 1.4147 & $102.9 \pm 44.2$ \\
    3mm.05 & 03:32:39.76 & -27:46:11.5 & $90.393 \pm 0.006$ & $2\rightarrow1$ & 1.5504 & 1001 & 1.5509 & $71.7 \pm 44.7$ \\
    3mm.06 & 03:32:39.90 & -27:47:15.1 & $110.038 \pm 0.005$ & $2\rightarrow1$ & 1.0951 & 8 & 1.0955 & $79.2 \pm 42.3$ \\
    3mm.07 & 03:32:43.53 & -27:46:39.4 & $93.558 \pm 0.008$ & $3\rightarrow2$ &  2.6961 & \nodata & \nodata & \nodata \\
    3mm.08 & 03:32:35.58 & -27:46:26.1 & $96.778 \pm 0.002$ & $2\rightarrow1$ & 1.3821 & 6415 & 1.3820 & $-0.1 \pm 40.5$ \\
    3mm.09 & 03:32:44.03 & -27:46:36.0 & $93.517 \pm 0.003$ & $3\rightarrow2$ & 2.6977$^{\dagger}$ & \nodata & \nodata & \nodata \\
    3mm.10 & 03:32:42.98 & -27:46:50.4 & $113.192 \pm 0.009$ & $2\rightarrow1$ & 1.0367 & 1011 & 1.0362$^{*}$ & $-53.7 \pm 46.6$ \\
    3mm.11 & 03:32:39.80 & -27:46:53.7 & $109.966 \pm 0.003$ & $2\rightarrow1$ & 1.0964 & 16 & 1.0965 & $19.8 \pm 40.8$ \\
    3mm.12 & 03:32:36.21 & -27:46:27.7 & $96.757 \pm 0.004$ & $3\rightarrow2$ & 2.5739 & 1124$^{\ddagger}$ & 2.5739$^{*}$ & $16.8 \pm 41.9$ \\
    3mm.13 & 03:32:35.56 & -27:47:04.3 & $100.209 \pm 0.006$ & $4\rightarrow3$ & 3.6008 & \nodata & \nodata & \nodata \\
    3mm.14 & 03:32:34.84 & -27:46:40.7 & $109.877 \pm 0.009$ & $2\rightarrow1$ & 1.0981 & 924 & 1.0981 & $15.0 \pm 46.9$ \\
    3mm.15 & 03:32:36.48 & -27:46:31.9 & $109.971 \pm 0.005$ & $2\rightarrow1$ & 1.0964 & 6870 & 1.0979 & $240.4 \pm 42.3$ \\
    3mm.16 & 03:32:39.92 & -27:46:07.4 & $100.503 \pm 0.004$ & $2\rightarrow1$ & 1.2938 & 925 & 1.2942 & $66.3 \pm 41.7$ \\
    \hline
    \multicolumn{9}{l}{\textbf{Notes.} $^{*}$Updated from \cite{Inami2017}, see \autoref{sec:redsh-conf}.} \\ %
    \multicolumn{9}{l}{$^{\dagger}$Additionally supported by matching absorption found in MUSE\#6941, at $z=2.695$, $0\farcs7$ to the north.} \\
    \multicolumn{9}{l}{$^{\ddagger}$Additional redshift for MUSE\#1124, which is cataloged as the foreground \OII-emitter at $z=1.098$ (see \autoref{fig:ASPECS12}).}
  \end{tabular}
\end{table*}

An extensive description of the line search is provided in
\cite{Gonzalez-LopezALP}.  In summary, three independent methods were combined
to search for CO lines in the ASPECS-LP band 3 data without any preselection;
\textsc{LineSeeker} \citep{Gonzalez-Lopez2017b}, \textsc{FindClump}
\citep{Decarli2014, Walter2016} and \textsc{MF3D} \citep{Pavesi2018}.  The
fidelity\footnote{The fidelity is defined as $F=1-P$, where $P$ is the
  probability of a line being produced by noise \citep{Gonzalez-LopezALP}.} of
these line-candidates was estimated from the ratio of the number of lines with
a negative and positive flux detected at a given S/N.  Lastly, the completeness
of the sample was estimated by ingesting simulated emission lines into the real
data cube.

In total, there are 16 emission line candidates for which the fidelity is
$\ge0.9$.  Statistical analysis shows that this sample is free from false
positives (the sum of their fidelities, based on the ALMA data alone, is 15.9;
\citealt{Gonzalez-LopezALP}).  These 16 sources form the primary, \emph{line
  search}-sample and are shown in \autoref{fig:HST-cutouts-CO-det}.  All these
candidates have a $\mathrm{S/N} \ge 6.4$.

For all sources in the primary sample, one or multiple potential counterpart
galaxies are visible in the deep \emph{HST} imaging shown in
\autoref{fig:HST-cutouts-CO-det}.  In order to confidently identify a single CO
emission line, an independent redshift measurement of the potential counterpart
measurement is needed.  Given the wealth of multi-wavelength photometry in the
HUDF, photometric redshifts can often already provide sufficient constraints to
discern between different rotation transitions of CO in the case of isolated
galaxies at redshifts $z \lesssim 3$.  However, complex systems of several
galaxies, or projected superpositions of independent galaxies at distinct
redshifts, can make redshift assignments more complicated.  Fortunately, the
integral-field spectroscopy from MUSE is ideally suited to disentangle spectral
features belonging to different galaxies, allowing us to confidently assign
redshifts to the CO emission lines.  The frequency of a CO line can correspond
to different rotational transitions, each with a unique associated redshift.
With the potential redshift solutions in hand, we systematically identify the
CO line candidates from the line search.  We provide a summary of the redshift
identifications here.  A detailed description of the individual sources and
their redshift identifications can be found in \autoref{sec:redsh-conf}, where
we also show the MUSE spectra for all sources (\autoref{fig:MUSE-spectra-1} --
\ref{fig:MUSE-spectra-4}).

First, we correlate the spatial position and potential redshifts of the CO
lines with known spectroscopic redshifts from MUSE \citep{Inami2017}.  From the
MUSE redshifts alone, we immediately identify most (11/16) of the CO lines with
the highest fidelity.  The brightest (ASPECS-LP.3mm.01) is a CO(3-2) emitter at
$z=2.54$, showing a wealth of UV absorption features.  The other 10 galaxies
are a diverse sample of CO(2-1) emitters spanning the redshift range over which
we are sensitive; $1.01 < z < 1.74$.  They show a variety of spectra at
different levels of S/N, covering a range of UV and optical absorption and
emission features.  Notably, \Oii\ is detected in all galaxies where it is
covered by MUSE, while \Neiii\ is detected in some of the higher S/N spectra.

Next, we extract MUSE spectra for the remaining five (5/16) sources without a
cataloged redshift and investigate their spectra for a redshift solution
matching the observed CO line.  We discover two new spectroscopic redshifts at
$z=2.54$ (ASPECS-LP.3mm.12) and $z=2.69$ (associated with ASPECS-LP.3mm.09)
confirming detections of CO(3-2), which were both not included in the catalog
of \cite{Inami2017} as their spectra are blended with foreground sources.  The
former in particular demonstrates the key use of MUSE in disentangling a
spatially overlapping system comprised of a foreground \OII\ emitter and a
faint background galaxy, which is detected at $\mathrm{S/N} > 4$ both via
cross-correlation with a $z\approx2.5$ spectral template and by stacking
absorption features (see \autoref{fig:ASPECS12}).  For ASPECS-LP.3mm.03 and
ASPECS-LP.3mm.07 we leverage the absence of spectral features (e.g., \OII,
\Lya), consistent with their faint magnitudes ($m_{\mathrm{F775W}} > 27$ mag)
and a redshift in the MUSE redshift desert, in combination with photometric
redshifts in the $z=2-3$ regime from the deep multi-wavelength data, to confirm
detections of CO(3-2).  Lastly, we find ASPECS-LP.3mm.13 being CO(4-3) at
$z=3.601$, based on the photometric redshifts suggesting $z\approx 3.5$ and the
absence of a lower redshift solution from the spectrum.  \Lyalpha\ is not
detected for this source, but we caution that at this redshift \Lya\ falls very
close to the \Oisky\ skyline.  Furthermore, given that the source potentially
contains significant amounts of dust, no \Lya\ emission may be expected at all.

In summary, we determine a redshift solution for all (16/16) candidates from
the line search.  Twelve are directly confirmed by MUSE spectroscopy, while the
remaining four are supported by their photometric redshifts and indirect
spectroscopic evidence.  We highlight that some of these counterparts are very
faint, even in the reddest \emph{HST} bands, and their identifications would
not have been possible without the exquisite depth of both the \emph{HST} and
MUSE data over the HUDF.  Similar objects would typically not have robust
photometric counterparts in areas of the sky with inferior coverage (let alone
have independent spectroscopic confirmation).

The identifications of the CO transitions, along with their MUSE counterparts,
are presented in \autoref{tab:CO-line-iden}.  We show the spatial extent of the
CO emission on top of the \emph{HST} images in
\autoref{fig:HST-cutouts-CO-det}.  The MUSE spectra for the individual sources
are shown in \autoref{fig:MUSE-spectra-1} -- \ref{fig:MUSE-spectra-4} and
discussed in \autoref{sec:redsh-conf}.

\subsection{Additional sources with MUSE redshift priors at $z < 2.9$}
\label{sec:addit-aspecs-sourc}
\begin{figure}[t]
  \includegraphics[width=\columnwidth]{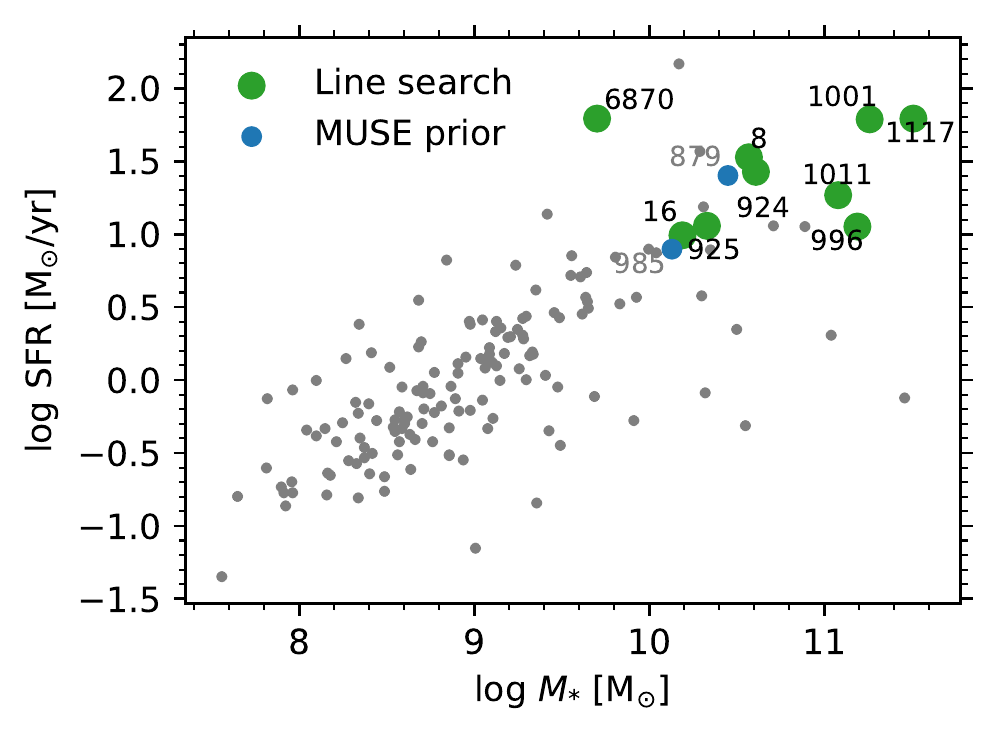}
  \caption{The stellar mass vs. SFR (from \textsc{Magphys}) of all galaxies
    with a MUSE redshift at $1.01 < z < 1.74$ in the ASPECS-LP footprint.
    Leveraging the MUSE redshift as prior, we find CO(2-1) signal in two
    additional galaxies (blue).  The numbers indicate the MUSE IDs of the
    sources.  The detections from the line-search (green;
    \autoref{sec:redsh-ident}) are also recovered in the prior-based search.
    By using the MUSE redshifts to search for CO at lower luminosities, we
    reveal molecular gas in most of the massive, star-forming galaxies at these
    redshifts. \label{fig:fit-muse-msfr}}
\end{figure}

\begin{figure*}[t]
  \includegraphics[width=\textwidth]{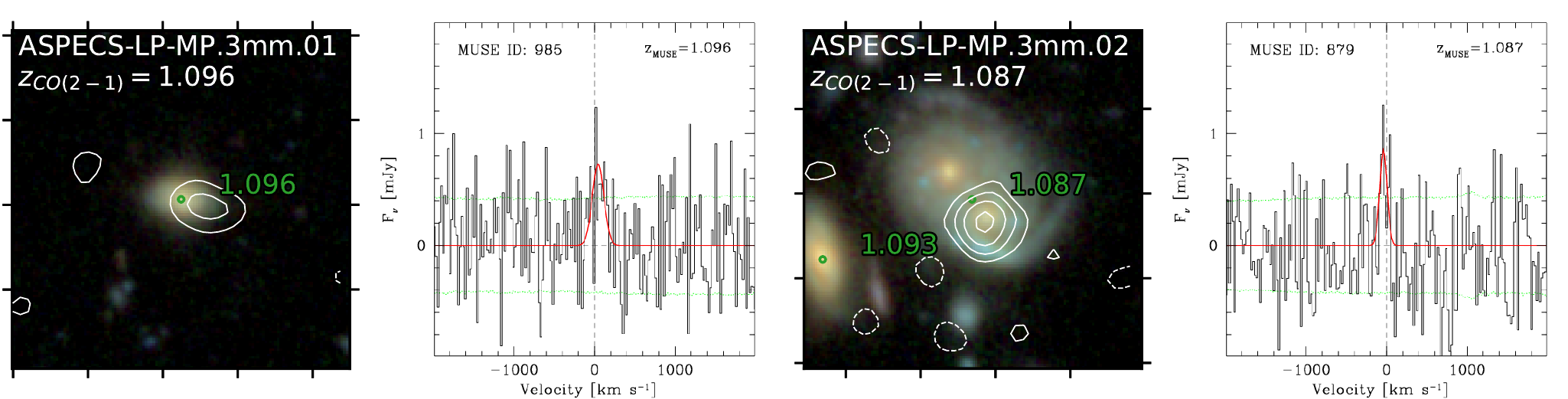}
  \caption{\emph{HST} cutouts (F160W, F125W, F105W) and CO(2-1) spectra for two
    additional CO line candidates, found through a MUSE redshift prior.  The CO
    contours are shown in white starting at $\pm2\sigma$ in steps of $1\sigma$.
    All other labelling in the cutouts is as in \autoref{fig:fit-muse-msfr}.
    In the spectra the velocity is given relative to the MUSE redshift.  The
    spectrum and best-fit Gaussian are shown in black and red, respectively.
    The local rms noise level is shown in green.\label{fig:spectra-prior}}
\end{figure*}

\begin{table*}[t]
  \centering
  \caption{ASPECS-LP CO(2-1) detected sources based on a spectroscopic redshift
    prior from MUSE.  (1) ASPECS-LP Muse Prior (MP) ID (2)-(3) Coordinates.
    (4) CO line Frequency.  (5) CO transition. (6) CO redshift.  (7) MUSE
    ID. (8) MUSE redshift. (9) Velocity offset between MUSE and ALMA
    ($\Delta v = (z_{\rm MUSE} - z_{\rm CO})/(1 + z_{\rm CO})$; after
    converting both to the same reference frame). \label{tab:CO-line-prior}}
  \begin{tabular}{ccccccccc}
    \hline\hline
    ID & R.A.    & Dec.   & $\nu_{\mathrm{CO}}$ & CO trans. & $z_{\mathrm{CO}}$ & MUSE ID & $z_{\mathrm{MUSE}}$ & $\Delta v$\\
     & (J2000) & (J2000) & (GHz) & ($J_{\rm up}\rightarrow J_{\rm low}$) & & & & km~s$^{-1}$\\
    (1) & (2) & (3) & (4) & (5) & (6) & (7) & (8) & (9) \\
    \hline
    MP.3mm.01 & 03:32:37.30 & -27:45:57.8 & $109.978 \pm 0.011$ & $2\rightarrow1$ & 1.0962 & 985 & 1.0959 & $-28.2 \pm 50.6$ \\
    MP.3mm.02 & 03:32:35.48 & -27:46:26.5 & $110.456 \pm 0.007$ & $2\rightarrow1$ & 1.0872 & 879 & 1.0874 & $55.8 \pm 44.3$ \\
    \hline
  \end{tabular}
\end{table*}
The CO-line detections from \cite{Gonzalez-LopezALP} are selected to have the
highest fidelity and are therefore the highest S/N ($\ge 6.4$) candidates over
the ASPECS-LP area.  In \autoref{fig:fit-muse-msfr}, we plot the stellar mass -
SFR relation for all MUSE sources at $1.01 < z < 1.74$, where we indicate all
the galaxies that have been detected in CO(2-1) in the line
search.\footnote{Note that we do not show the MUSE source associated with
  ASPECS-LP.3mm.08 and the two MUSE sources that are severely blended with
  ASPECS-LP.3mm.12 and the galaxy north of ASPECS-LP.3mm.09 on the plot.}
There are several galaxies in the field with properties similar to the
ASPECS-LP galaxies that are not detected in the line search.  This raises the
question: Why are these galaxies not detected?  Given their physical
properties, we may expect some of these galaxies to harbor molecular gas and
therefore to have CO signal in the ASPECS-LP cube.  The reason that we did not
detect these sources in the line search may, therefore, simply be due to the
fact that they are present at lower S/N, which puts them in the regime where
the decreasing fidelity makes it challenging to identify them among the
spurious sources.

However, the physical properties of the galaxies themselves provide an extra
piece of information that can guide us in detecting CO for these sources.  In
particular, we can use the spectroscopic redshifts from MUSE to obtain a
measurement of the CO flux for each source, either identifying them at lower
S/N, or putting an upper limit on their molecular gas mass.  We aim at the CO
transitions covered at $z<2.9$, where the features in the MUSE spectrum
typically provide a systemic redshift.  At higher redshift the main spectral
feature used to identify redshifts is often \Lya, which can be offset from the
systemic redshift by a few hundred km~s$^{-1}$ \citep[e.g.,][]{Shapley2003,
  Rakic2011, Verhamme2018}.

We extract a single-pixel spectrum from the 3\arcsec\ tapered cube at the
position of each MUSE source in the redshift range, after correcting for the
astrometric offset (\autoref{sec:alma}).  We then fit the lines with a Gaussian
curve, using a custom-made Bayesian Markov chain Monte Carlo routine with the
following priors:
\begin{itemize}
\item \emph{line peak velocity}: a Gaussian distribution centered at
  $\Delta v = 0$ (based on the MUSE redshift) and $\sigma = 100$ km~s$^{-1}$
  (the MUSE spectral resolution).
\item \emph{line width}: a Maxwellian distribution with a width of 100
  km~s$^{-1}$.
\item \emph{line flux}: a Gaussian distribution centered at zero, with
  $\sigma = 0.5$ Jy km~s$^{-1}$, allowing both positive and negative line
  fluxes to be fitted.
\end{itemize}
We choose a strong prior on the velocity difference, as we only search for
lines at the exact MUSE redshift.  The Gaussian prior on the line flux is
important to estimate the fidelity of our measurements, allowing an unbiased
comparison of positive versus negative line fluxes (see
\citealt{Gonzalez-LopezALP} for details).  The Maxwellian prior is chosen
because it is bound to produce positive values of the line-width, depends on a
single scale parameter and has a non-null tail at very large line widths.  The
uncertainties are computed from the 16th and 84th percentiles of the posterior
distributions of each parameter.

As narrow lines are more easily caused by noise in the cube
\citep{Gonzalez-LopezALP}, we rerun the fit with a broader prior on the line
width of 200~km~s$^{-1}$.  We also independently fit the spectrum with a
uniform prior over $\pm 1$~GHz around the MUSE-redshift.  We select only the
sources in which the same feature was recovered with $\mathrm{S/N}>3$ in all
three fits.  In order to select a sample that is as pure as possible, we select
only the objects that have a velocity offset of $<80$~km~s$^{-1}$ from the MUSE
systemic redshift ($\approx\times2$ the typical uncertainty on the MUSE
redshift).  In addition, we only keep objects with a line width of
$>100$~km~s$^{-1}$, to avoid including spurious narrow lines.  We note that,
while these cuts potentially remove other sources that are detected at lower
S/N, we do not attempt to be complete.  Rather, we aim to have the prior-based
sample as clean as possible.

The prior-based search reveals two additional sources detected in CO(2-1) with
a $\mathrm{S/N} > 3$.  Both sources lie within the area in which the
sensitivity is $>40\%$ of the primary beam peak sensitivity.  We show the
\emph{HST} cutouts with the CO spectra of these sources in
\autoref{fig:spectra-prior}, ordered by S/N.  ASPECS-LP-MP.3mm.02 is the
foreground spiral galaxy of ASPECS-LP.3mm.08.  This source was already found in
the ASPECS-Pilot \citep[][see \autoref{sec:redsh-conf}]{Decarli2016}.

Because the molecular gas mass is to first order correlated with the SFR, we
expect to detect CO in the galaxies with the highest SFRs at a given redshift.
Sorting all the galaxies by their SFR indeed reveals a clear correlation
between the SFR and the S/N in CO, suggesting there are additional sources in
the ASPECS-LP datacube at lower S/N.  This can also be clearly seen from
\autoref{fig:fit-muse-msfr}, where our stringent sample of prior based sources
all lie at log~SFR[\Msun~yr$^{-1}$]~$>0.5$.  Qualitatively, it becomes clear
that the ASPECS-LP is sensitive enough to detect molecular gas in most massive
main sequence galaxies at $1.01 < z < 1.74$ (a quantitative discussion of the
detection fraction for the full sample is provided in
\autoref{sec:discussion}).  For many galaxies, the reason these are not
unveiled in the line search may simply be because their lower CO luminosity
and/or smaller line-width puts them below the conservative S/N threshold we
adopt in the line search.  Using the MUSE redshifts as prior information, it is
possible to unveil their molecular gas reservoirs at lower S/N.

\subsection{Full sample redshift distribution}
\label{sec:combined-sample}
\begin{figure}[t]
  \includegraphics[width=\columnwidth]{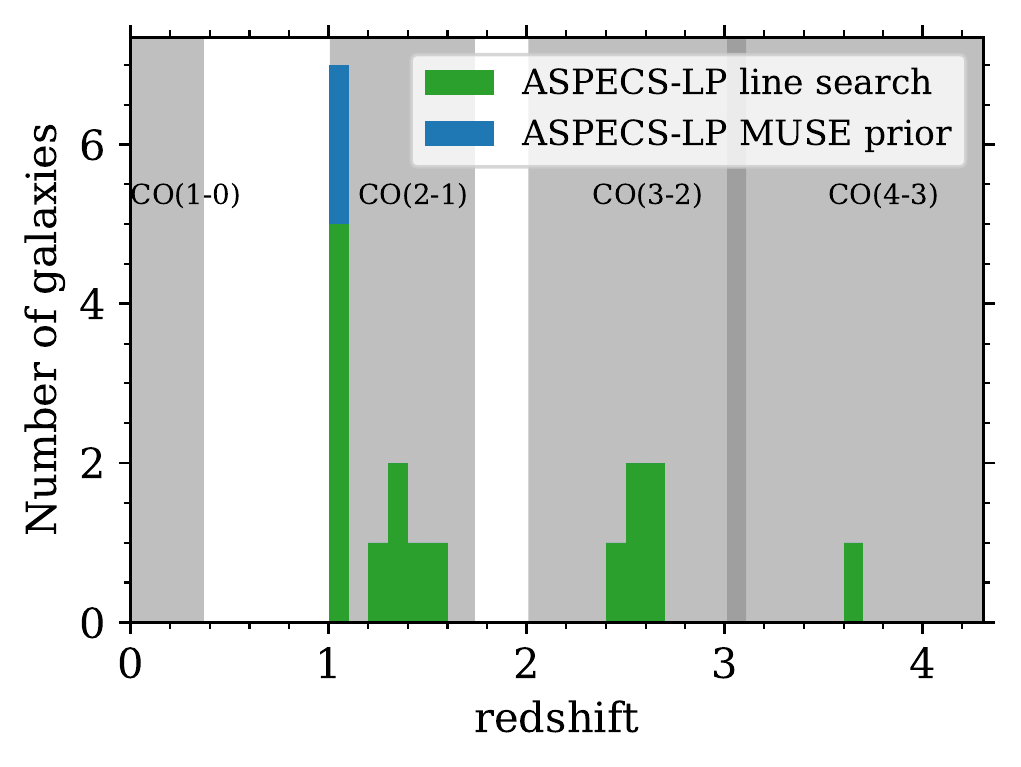}
  \caption{Redshift distribution of the ASPECS-LP CO detected sources, which
    all have a \emph{HST} counterpart.  We show both the detections from the
    line search (\autoref{sec:methods}) as well as the MUSE prior based
    galaxies (\autoref{sec:addit-aspecs-sourc}).  The gray shading indicates
    the redshift ranges over which we can detect different CO
    transitions. \label{fig:hist-redshift}}
\end{figure}

The full ASPECS-LP CO line sample consists of 18 galaxies with a CO detection
in the HUDF; 16 detections without preselection and 2 MUSE redshift prior based
detections.  These galaxies span a range of redshifts between $1 < z < 4$.  The
lowest redshift galaxy is detected in CO(2-1) at $z=1.04$, while the highest
redshift galaxy is detected (without prior) in CO(4-3) at $z=3.60$.  We show a
histogram of the redshifts of the line-search and prior-based detections in
\autoref{fig:hist-redshift}.

Twelve sources are detected in CO(2-1) at $1.01 < z < 1.74$, where the
combination of molecular line sensitivity and survey volume are optimal.  Most
prominently, we detect five galaxies at the same redshift of $z \approx 1.1$.
These galaxies are all part of an overdensity of galaxies in the HUDF at
$z = 1.096$, visible in \autoref{fig:MUSE-zdist}.

Five sources are detected in CO(3-2) at $2.01 < z < 3.11$, including the
brightest CO emitter in the field at $z=2.54$ (ASPECS-LP.3mm.01; see also
\citealt{Decarli2016}) and a pair of galaxies (ASPECS-LP.3mm.07 and \#9) at
$z\approx2.697$ (see \autoref{sec:redsh-ident}).  All five CO(3-2) sources are
detected in 1~mm dust continuum \citep{Aravena2016, Dunlop2017} with flux
densities below 1~mJy.  However, only one of these sources (ASPECS-LP.3mm.01)
previously had a spectroscopic redshift \citep{Walter2016, Inami2017}.

\section{Physical properties}
\label{sec:deriv-phys-prop}

\subsection{Star formation rates from \textsc{magphys} \& \OII}
\label{sec:sfrs}

\begin{table*}[t]
  \centering
  \caption{Physical properties of the ASPECS-LP detected sources from the line
    search and the MUSE prior-based search with formal uncertainties. (1)
    ASPECS-LP ID number.  (2) Source redshift. (3) Stellar mass ($M_{*}$). (4)
    Star formation rate (SFR). (5) Visual attenuation ($A_{V}$). (6)--(7) X-ray
    classification as active galactic nucleus (AGN) or other X-ray source (X)
    from \cite{Luo2017} and corresponding X-ray ID
    (XID). \label{tab:derived-properties}}
  \begin{tabular}{ccccccc}
    \hline
    \hline
    ID &  $z$ & $\log M_{*,\mathrm{SED}}$ & SFR$_{\mathrm{SED}}$ & $A_{V, \mathrm{SED}}$ & X-ray & XID\\
    &                     & (\solarmass)              &    (\solarmass~yr$^{-1}$) & (mag) & & \\
    (1)              & (2)                    & (3) & (4) & (5) & (6)  & (7) \\
    \hline
    ASPECS-LP.3mm.01 & 2.5436 & $10.4_{-0.0}^{+0.0}$ & $233_{-0}^{+0}$ & $2.7_{-0.0}^{+0.0}$ & AGN & 718 \\
    ASPECS-LP.3mm.02 & 1.3167 & $11.2_{-0.0}^{+0.0}$ & $11_{-0}^{+2}$ & $1.7_{-0.0}^{+0.1}$ &  &  \\
    ASPECS-LP.3mm.03 & 2.4534 & $10.7_{-0.1}^{+0.1}$ & $68_{-20}^{+19}$ & $3.1_{-0.3}^{+0.1}$ &  &  \\
    ASPECS-LP.3mm.04 & 1.4140 & $11.3_{-0.0}^{+0.0}$ & $61_{-12}^{+3}$ & $2.9_{-0.0}^{+0.1}$ &  &  \\
    ASPECS-LP.3mm.05 & 1.5504 & $11.5_{-0.0}^{+0.0}$ & $62_{-19}^{+5}$ & $2.3_{-0.3}^{+0.1}$ & AGN & 748 \\
    ASPECS-LP.3mm.06 & 1.0951 & $10.6_{-0.0}^{+0.0}$ & $34_{-0}^{+0}$ & $0.8_{-0.0}^{+0.0}$ & X & 749 \\
    ASPECS-LP.3mm.07 & 2.6961 & $11.1_{-0.1}^{+0.1}$ & $187_{-16}^{+35}$ & $3.2_{-0.1}^{+0.1}$ &  &  \\
    ASPECS-LP.3mm.08 & 1.3821 & $10.7_{-0.0}^{+0.0}$ & $35_{-5}^{+8}$ & $0.9_{-0.1}^{+0.1}$ &  &  \\
    ASPECS-LP.3mm.09 & 2.6977 & $11.1_{-0.0}^{+0.1}$ & $318_{-35}^{+35}$ & $3.6_{-0.1}^{+0.1}$ & AGN & 805 \\
    ASPECS-LP.3mm.10 & 1.0367 & $11.1_{-0.1}^{+0.0}$ & $18_{-1}^{+1}$ & $3.0_{-0.1}^{+0.0}$ &  &  \\
    ASPECS-LP.3mm.11 & 1.0964 & $10.2_{-0.0}^{+0.0}$ & $10_{-1}^{+0}$ & $0.8_{-0.1}^{+0.0}$ &  &  \\
    ASPECS-LP.3mm.12 & 2.5739 & $10.6_{-0.1}^{+0.0}$ & $31_{-3}^{+18}$ & $0.8_{-0.1}^{+0.2}$ & AGN & 680 \\
    ASPECS-LP.3mm.13 & 3.6008 & $9.8_{-0.1}^{+0.1}$ & $41_{-9}^{+15}$ & $1.4_{-0.2}^{+0.3}$ &  &  \\
    ASPECS-LP.3mm.14 & 1.0981 & $10.6_{-0.1}^{+0.1}$ & $27_{-4}^{+1}$ & $1.6_{-0.2}^{+0.0}$ &  &  \\
    ASPECS-LP.3mm.15 & 1.0964 & $9.7_{-0.0}^{+0.3}$ & $62_{-4}^{+0}$ & $2.9_{-0.0}^{+0.0}$ & AGN & 689 \\
    ASPECS-LP.3mm.16 & 1.2938 & $10.3_{-0.0}^{+0.1}$ & $11_{-3}^{+1}$ & $0.5_{-0.2}^{+0.1}$ &  &  \\
    \hline
    ASPECS-LP-MP.3mm.01 & 1.0959 & $10.1_{-0.0}^{+0.1}$ & $8_{-2}^{+3}$ & $1.3_{-0.2}^{+0.2}$ &  &  \\
    ASPECS-LP-MP.3mm.02 & 1.0874 & $10.4_{-0.0}^{+0.0}$ & $25_{-0}^{+0}$ & $1.0_{-0.0}^{+0.0}$ & X & 661 \\
    \hline
  \end{tabular}
\end{table*}

\begin{table*}[t]
  \centering
  \caption{Emission line flux measurements and derived unobscured SFR and
    metallicity for the ASPECS-LP line-search and prior-based sources at
    $z < 1.5$ with $\mathrm {S/N}(\OII) >3$.  (1) ASPECS-LP.3mm ID number. (2)
    MUSE ID (3) MUSE redshift. (4) \Oiisum\ flux ($\mathrm{S/N} > 3$). (5)
    \Neiii\ flux (upper limits are reported if $\mathrm{S/N} < 3$).  (6)
    SFR(\Oii) without correction for dust attenuation. (7) Metallicity from
    \NeIII/\OII\ based on
    \cite{Maiolino2008}. \label{tab:emission-line-properties}}
  \begin{tabular}{ccccccc}
    \hline\hline
    ID & MUSE ID &  $z_{\mathrm{MUSE}}$ &    $F_{\Oiisum}$ & $F_{\Neiii}$ & SFR$_{\OII}^{\rm no~dust}$ & $Z_{\NeIII/\OII, \mathrm{M08}}$ \\
    & & & ($\times 10^{-20}\mathrm{erg\,s^{-1}\,cm^{-2}}$) & ($\times 10^{-20}\mathrm{erg\,s^{-1}\,cm^{-2}}$) & (\solarmass\,yr$^{-1}$) & (12 + log(O/H)) \\
    (1) & (2) & (3) & (4) & (5) & (6) & (7) \\
    \hline
    3mm.06 & 8 & 1.0955 & $111.4\pm1.4$ & $1.9\pm0.4$ & $3.59\pm0.05$ & $9.05 \pm 0.08$ \\
    3mm.11 & 16 & 1.0965 & $24.4\pm0.3$ & $0.9\pm0.1$ & $0.79\pm0.01$ & $8.78 \pm 0.06$ \\
    3mm.14 & 924 & 1.0981 & $53.6\pm1.6$ & $2.4\pm0.4$ & $1.74\pm0.05$ & $8.70 \pm 0.07$ \\
    3mm.15 & 6870 & 1.0979 & $13.8\pm0.4$ & $<0.2\pm0.1$ & \nodata & \nodata \\
    3mm.16 & 925 & 1.2942 & $67.0\pm4.0$ & $<1.9\pm0.8$ & $3.26\pm0.20$ & $>8.79 \pm 0.17$ \\
    \hline
    MP.3mm.01 & 985 & 1.0959 & $17.8\pm1.5$ & $<0.6\pm0.5$ & $0.57\pm0.05$ & $>8.56 \pm 0.29$ \\
    MP.3mm.02 & 879 & 1.0874 & $245.9\pm1.1$ & $11.5\pm0.6$ & $7.78\pm0.03$ & $8.73 \pm 0.02$ \\
    \hline
    \multicolumn{7}{l}{\textbf{Notes.} We do not compute a SFR(\OII) or metallicity for the X-ray detected AGN (3mm.15).}
\end{tabular}
\end{table*}

For all the CO detected sources, we derive the SFR (and $M_{*}$ and $A_{V}$)
from the UV-FIR data (including 24\micron--160\micron\ and ASPECS-LP 1.2~mm and
3.0~mm) using \textsc{Magphys} (see \autoref{sec:multi-wavel-data}), which are
provided in \autoref{tab:derived-properties}.  The full SED fits are shown in
\autoref{fig:fig_sed} and \ref{fig:fig_sed2}.

For the $1 < z < 1.5$ subsample, we have access to the \Oii-doublet.  We derive
SFRs from \Oii\ following \cite{Kewley2004}, adopting a \cite{Chabrier2003}
IMF.  The observed \OII\ luminosity gives a measurement of the unobscured SFR,
which can be compared to the total SFR (including the FIR) to derive the
fraction of obscured star formation.  For that reason, we not apply a dust
correction when calculating the SFR(\OII).

The derived SFR(\Oii) is dependent on the oxygen abundance.  We have access to
the oxygen abundance directly for some of the sources and can also make an
estimate through the mass-metallicity relation \citep[e.g.,][]{Zahid2014}.
However, because of the additional uncertainties in the calibrations for the
oxygen abundance, we instead adopt an average \Oii/\Ha\ ratio of unity, given
that all our sources are massive and hence expected to have high oxygen
abundance \logOH\ $\sim 8.8$, where $\OII/\Ha=1.0$ \citep[e.g.,][]{Kewley2004}.
For all galaxies with S/N(\Oii)~$>3$, excluding the X-ray AGN, the \Oiisum\
line flux measurements and SFRs are presented in
\autoref{tab:emission-line-properties}.

\subsection{Metallicities}
\label{sec:metallicities}
It is well known that the gas-phase metallicity of galaxies is correlated with
their stellar mass, with more massive galaxies having higher metallicities on
average \citep[e.g.,][]{Tremonti2004, Maiolino2008, Mannucci2010, Zahid2014}.
For the $1.0 < z < 1.42$ sub-sample, we have access to \Neiii\ which allows us
to derive a metallicity from \Neiii/\Oii.  We follow the relation as presented
by \cite{Maiolino2008}, who calibrated the \NeIII/\OII\ line ratio against
metallicities inferred from the direct $T_{e}$ method (at low metallicity;
$\logOH < 8.35$) and theoretical models from \cite{Kewley2002} (at high
metallicity, mainly relevant for this paper; $\logOH > 8.35$).  Since the
wavelengths of \Neiii\ and \Oii\ are close, this ratio is practically
insensitive to dust attenuation.  The physical underpinning lies in the fact
that the ratio of the low-ionization \OII\ and high-ionization \NeIII\ lines is
a solid tracer of the shape of the ionization field, given that neon closely
tracks the oxygen abundance \cite[e.g.,][]{Ali1991, Levesque2014, Feltre2018}.
As the ionization parameter decreases with increasing stellar metallicity
\citep{Dopita2006a, Dopita2006b} and the metallicity of the young ionizing
stars and their birth clouds is correlated, the ratio of \Neiii/\Oii\ is a
reasonable gas-phase metallicity diagnostic, albeit indirect, with significant
scatter \citep{Nagao2006, Maiolino2008} and sensitive to model assumptions
\citep[e.g.,][]{Levesque2014}.  If an AGN contributes significantly to the
ionizing spectrum, the emission lines may no longer only trace the properties
associated with massive star formation.  For this reason, we exclude the
sources with an X-ray AGN from the analysis of the metallicity.

We report the \NeIII\ flux measurements and \NeIII/\OII\ metallicities in
\autoref{tab:emission-line-properties}.  The solar metallicity is
$\logOH = 8.76 \pm 0.07$ \citep{Caffau2011}.

\subsection{Molecular gas properties}
\label{sec:molec-gas-prop}
\begin{table*}[t]
  \centering
  \caption{Molecular gas properties of the ASPECS-LP line-search and
    prior-based sources with formal uncertainties.  The CO full-width at half
    maximum (FWHM) and line fluxes are taken from \cite{Gonzalez-LopezALP}.
    (1) ASPECS-LP ID number. (2) CO redshift. (3) Upper level of CO
    transition. (4) CO line FWHM. (5) Integrated line flux. (6) Line luminosity
    (7) CO(1-0) line luminosity assuming \cite{Daddi2015} excitation
    (\autoref{sec:molec-gas-prop}). (8) Molecular gas mass assuming
    $\aco = 3.6$~K~(km~s$^{-1}$~pc$^{2}$)$^{-1}$. (9) Molecular-to-stellar
    mass ratio, $\MH/M_{*}$. (10) Depletion time, $\tdepl = \MH / \mathrm{SFR}$.
    \label{tab:co-properties}}
  \begin{tabular}{cccccccccc}
    \hline
    \hline
    ID &  $z_{\rm CO}$ & $J_{\rm up}$& FWHM & $F_{\rm line}$ & $L'_{\rm line}$ & $L'_{\rm CO(1-0)}$ & $\MH$ & $\MH/M_{*}$ & $t_{\rm depl}$ \\
    &      & & (km~s$^{-1}$) & (Jy~km~s$^{-1}$)   & \multicolumn{2}{c}{($\times 10^{9}$~K~km~s$^{-1}$~pc$^{2}$)} & ($\times 10^{10}$~\Msun)   &  & (Gyr)  \\
    (1)              & (2)                    & (3) & (4) & (5) & (6)    & (7)  & (8) & (9) & (10)\\
    \hline
    3mm.01 & 2.5436 & 3 & $517 \pm 21$ & $1.02 \pm 0.04$ & $33.9 \pm 1.3$ & $80.8 \pm 13.8$ & $29.1 \pm 5.0$ & $12.1 \pm 2.1$ & $1.2 \pm 0.2$ \\
    3mm.02 & 1.3167 & 2 & $277 \pm 26$ & $0.47 \pm 0.04$ & $10.7 \pm 0.9$ & $14.1 \pm 2.1$ & $5.1 \pm 0.7$ & $0.3 \pm 0.1$ & $4.5 \pm 0.8$ \\
    3mm.03 & 2.4534 & 3 & $368 \pm 37$ & $0.41 \pm 0.04$ & $12.8 \pm 1.3$ & $30.5 \pm 5.9$ & $11.0 \pm 2.1$ & $2.2 \pm 0.6$ & $1.6 \pm 0.6$ \\
    3mm.04 & 1.4140 & 2 & $498 \pm 47$ & $0.89 \pm 0.07$ & $23.2 \pm 1.8$ & $30.5 \pm 4.3$ & $11.0 \pm 1.6$ & $0.6 \pm 0.1$ & $1.8 \pm 0.3$ \\
    3mm.05 & 1.5504 & 2 & $617 \pm 58$ & $0.66 \pm 0.06$ & $20.4 \pm 1.9$ & $26.9 \pm 4.0$ & $9.7 \pm 1.4$ & $0.3 \pm 0.1$ & $1.6 \pm 0.4$ \\
    3mm.06 & 1.0951 & 2 & $307 \pm 33$ & $0.48 \pm 0.06$ & $7.7 \pm 1.0$ & $10.1 \pm 1.7$ & $3.6 \pm 0.6$ & $1.0 \pm 0.2$ & $1.1 \pm 0.2$ \\
    3mm.07 & 2.6961 & 3 & $609 \pm 73$ & $0.76 \pm 0.09$ & $27.9 \pm 3.3$ & $66.5 \pm 13.6$ & $23.9 \pm 4.9$ & $2.0 \pm 0.5$ & $1.3 \pm 0.3$ \\
    3mm.08 & 1.3821 & 2 & $50 \pm 8$ & $0.16 \pm 0.03$ & $4.0 \pm 0.7$ & $5.3 \pm 1.2$ & $1.9 \pm 0.4$ & $0.4 \pm 0.1$ & $0.5 \pm 0.2$ \\
    3mm.09 & 2.6977 & 3 & $174 \pm 17$ & $0.40 \pm 0.04$ & $14.7 \pm 1.5$ & $35.0 \pm 6.8$ & $12.6 \pm 2.5$ & $1.0 \pm 0.2$ & $0.4 \pm 0.1$ \\
    3mm.10 & 1.0367 & 2 & $460 \pm 49$ & $0.59 \pm 0.07$ & $8.5 \pm 1.0$ & $11.1 \pm 1.9$ & $4.0 \pm 0.7$ & $0.3 \pm 0.1$ & $2.2 \pm 0.4$ \\
    3mm.11 & 1.0964 & 2 & $40 \pm 12$ & $0.16 \pm 0.03$ & $2.6 \pm 0.5$ & $3.4 \pm 0.7$ & $1.2 \pm 0.3$ & $0.8 \pm 0.2$ & $1.2 \pm 0.3$ \\
    3mm.12 & 2.5739 & 3 & $251 \pm 40$ & $0.14 \pm 0.02$ & $4.8 \pm 0.7$ & $11.3 \pm 2.5$ & $4.1 \pm 0.9$ & $0.9 \pm 0.2$ & $1.3 \pm 0.5$ \\
    3mm.13 & 3.6008 & 4 & $360 \pm 49$ & $0.13 \pm 0.02$ & $4.3 \pm 0.7$ & $13.9 \pm 3.4$ & $5.0 \pm 1.2$ & $8.8 \pm 2.8$ & $1.2 \pm 0.5$ \\
    3mm.14 & 1.0981 & 2 & $355 \pm 52$ & $0.35 \pm 0.05$ & $5.6 \pm 0.8$ & $7.4 \pm 1.4$ & $2.7 \pm 0.5$ & $0.7 \pm 0.1$ & $1.0 \pm 0.2$ \\
    3mm.15 & 1.0964 & 2 & $260 \pm 39$ & $0.21 \pm 0.03$ & $3.4 \pm 0.5$ & $4.4 \pm 0.8$ & $1.6 \pm 0.3$ & $3.2 \pm 1.1$ & $0.3 \pm 0.1$ \\
    3mm.16 & 1.2938 & 2 & $125 \pm 28$ & $0.08 \pm 0.01$ & $1.8 \pm 0.2$ & $2.3 \pm 0.4$ & $0.8 \pm 0.1$ & $0.4 \pm 0.1$ & $0.7 \pm 0.2$ \\
    \hline
    MP.3mm.01 & 1.0962 & 2 & $169 \pm 21$ & $0.13 \pm 0.03$ & $2.1 \pm 0.5$ & $2.8 \pm 0.7$ & $1.0 \pm 0.2$ & $0.7 \pm 0.2$ & $1.3 \pm 0.5$ \\
    MP.3mm.02 & 1.0872 & 2 & $107 \pm 30$ & $0.10 \pm 0.03$ & $1.6 \pm 0.4$ & $2.0 \pm 0.6$ & $0.7 \pm 0.2$ & $0.3 \pm 0.1$ & $0.3 \pm 0.1$ \\
    \hline
  \end{tabular}
\end{table*}

The derivation of the molecular gas properties of our sources is detailed in
\cite{AravenaALP}.  For reference, we provide a brief summary here.

We convert the observed CO($J\rightarrow J-1$) flux to a molecular gas mass
(\MH) using the relations from \cite{Carilli2013}.  To convert the higher order
CO transitions to CO(1-0), we need to know the excitation dependent intensity
ratio between the CO lines, $r_{J1}$.  We use the excitation ladder as
estimated by \cite{Daddi2015} for galaxies on the MS, where
$r_{21} = 0.76 \pm 0.09$, $r_{31} = 0.42 \pm 0.07$ and $r_{41} = 0.31 \pm 0.06$
(see also \citealt{Decarli2016}).  To subsequently convert the CO(1-0)
luminosity to \MH, we use an
$\aco = 3.6$~\solarmass~(K~km~s$^{-1}$~pc$^2$)$^{-1}$, appropriate for
star-forming galaxies (\citealt{Daddi2010}; see \citealt{Bolatto2013} for a
review).  This choice of \aco\ is supported by our finding that the ASPECS-LP
sources are mostly on the MS and have \mbox{(near-)solar} metallicity (see
\autoref{sec:metallicities-at-1.0}).

With these conversions in mind, the molecular gas mass and derived quantities
we report here can easily be rescaled to different assumptions following:
$\MH/\Msun = \left(\alpha_{\mathrm{CO}}/r_{J1}\right)
L'_{\mathrm{CO}(J\rightarrow J-1)}/$(K~km~s$^{-1}$~pc$^2$).

\section{Results: Global sample properties}
\label{sec:results}
In this section we discuss the physical properties of all the ASPECS-LP sources
that were found in the line search (without preselection) and based on a MUSE
redshift prior.  Since the sensitivity of ASPECS-LP varies with redshift, we
discuss the galaxies detected in different CO transitions separately.  In terms
of the demographics of the ASPECS-LP detections, we focus on CO(2-1) and
CO(3-2), where we have the most detections.

\subsection{Stellar mass and SFR distributions}
\label{sec:stellar-mass-sfr}
\begin{figure*}[t]
  \includegraphics[width=0.5\textwidth]{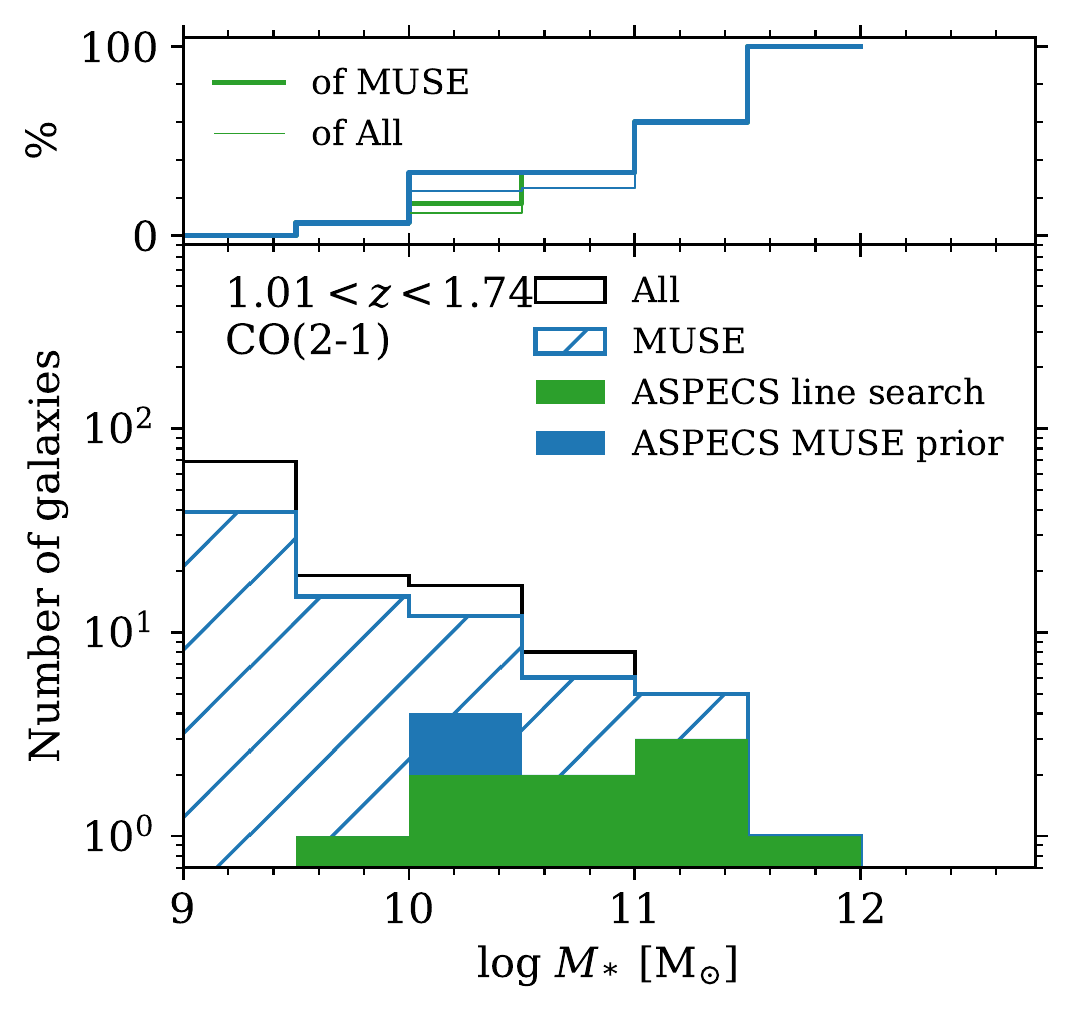}
  \includegraphics[width=0.5\textwidth]{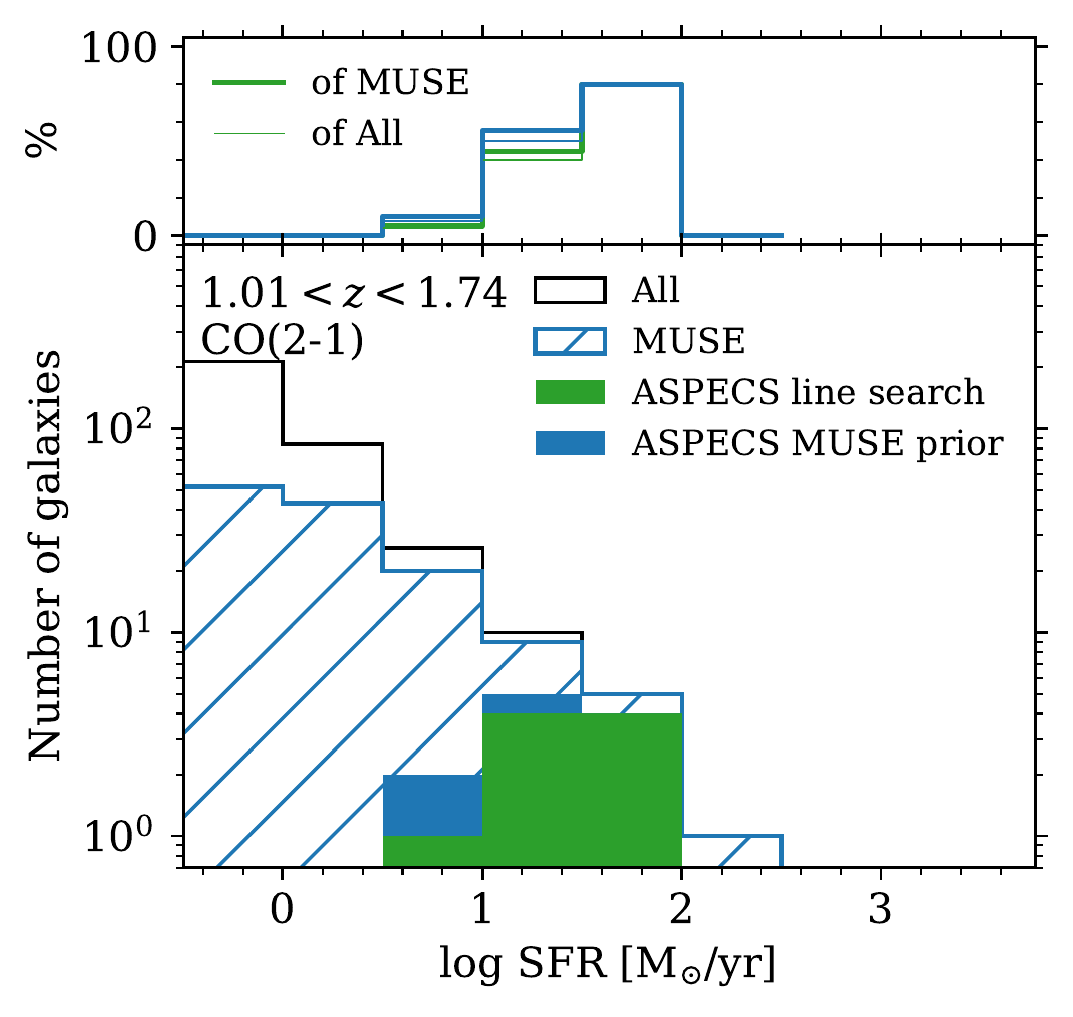}
  \includegraphics[width=0.5\textwidth]{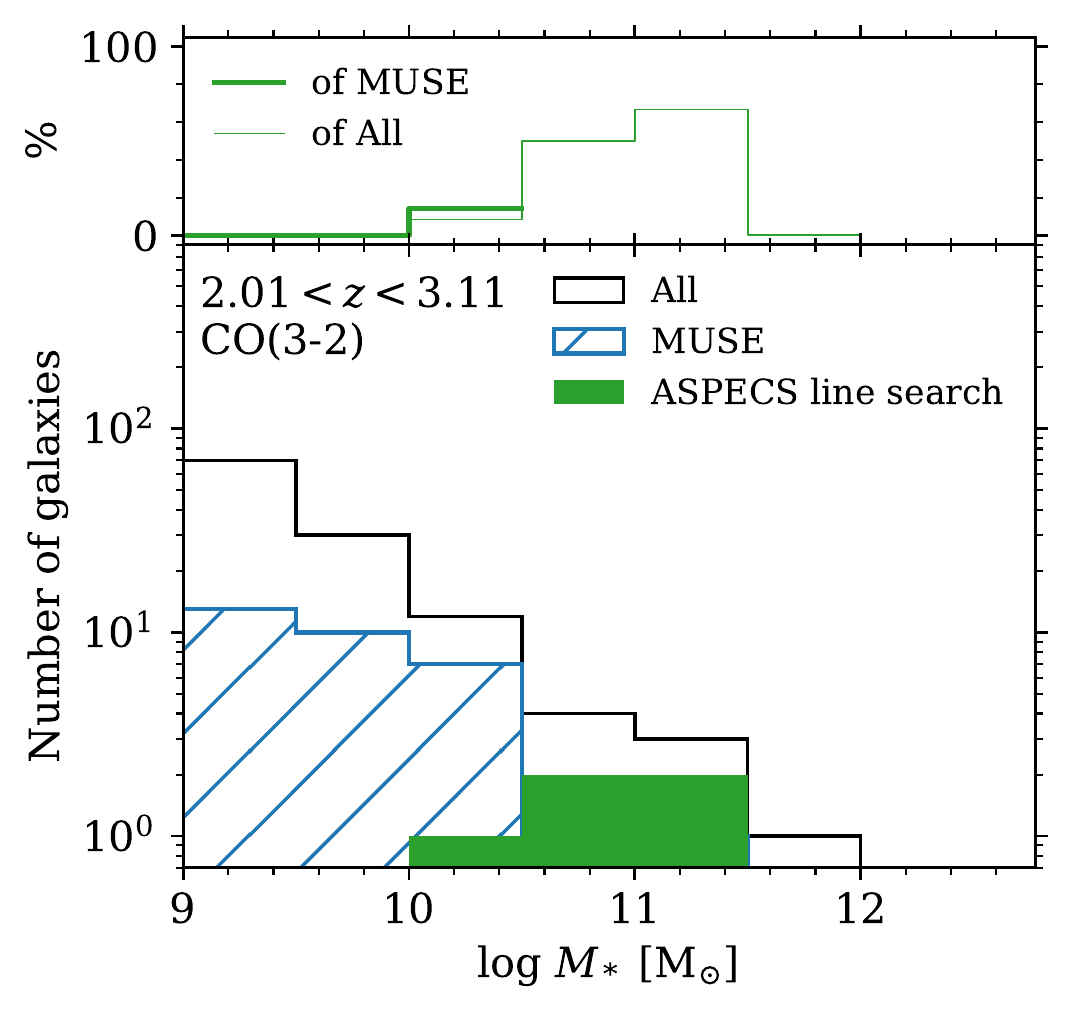}
  \includegraphics[width=0.5\textwidth]{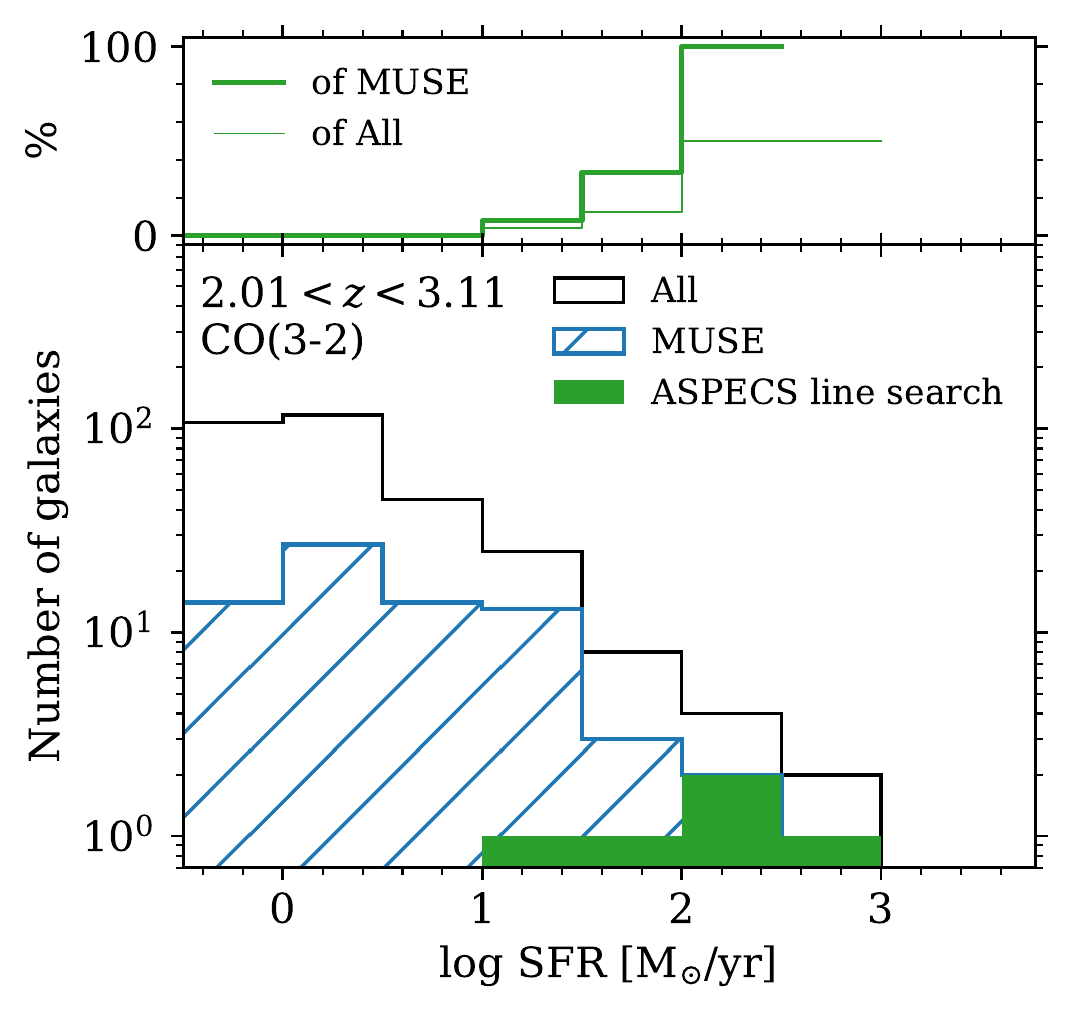}
  \caption{Histograms of the stellar mass ($M_{*}$, \textbf{left}) and star
    formation rate (SFR, \textbf{right}) of the ASPECS-LP detected galaxies, in
    comparison to all galaxies with MUSE redshifts and our extended photometric
    redshift catalog, in the indicated redshift range.  We only show the range
    relevant to the ASPECS-LP detections: $M_{*} > 10^{9}$~\Msun\ and
    SFR~$> 0.3$~\Msun/yr.  \textbf{Top:} CO(2-1) detected sources at
    $1.01 < z < 1.74$. \textbf{Bottom:} CO(3-2) detected sources at
    $2.01 < z < 3.11$.  In each of the four panels, the detected fraction in
    both reference catalogs is shown on top (no line is drawn if the catalog
    does not contain any objects in that bin).  With the ASPECS-LP, we detect
    approximately 40\% of (50\%) of all galaxies at $M_{*} > 10^{10}$~\Msun
    ($>10^{10.5}$~\Msun) at $1.0 < z < 1.7$ ($2.0 < z < 3.1$), respectively.
    In the same redshift bins, we detect approximately 60\% (30\%) of all
    galaxies with SFR~$>10$~\Msun/yr
    ($>30$~\Msun/yr). \label{fig:hist-mstar-sfr}}
\end{figure*}

The majority of the detections consist of CO(2-1) and CO(3-2), at $1 < z < 2$
and $2 < z < 3$, respectively.  A key question is in what part of the galaxy
population we detect the largest gas-reservoirs at these redshifts.

We show histograms of the stellar masses and SFRs for the sources detected in
CO(2-1) and CO(3-2) in \autoref{fig:hist-mstar-sfr}.  We compare these to the
distribution of all galaxies in the field that have a spectroscopic redshift
from MUSE and our extended (photometric) catalog of all other galaxies.  In the
top part of each panel we show the percentage of galaxies we detect in ASPECS,
compared to the number of galaxies in reference catalogs.

We focus first on the SFRs, shown in the right panels of
\autoref{fig:hist-mstar-sfr}.  The galaxies in which we detect molecular gas
are the galaxies with the highest SFRs and the detection fraction increases
with SFR.  This is expected as molecular gas is a prerequisite for star
formation and the most highly star-forming galaxies are thought to host the
most massive gas reservoirs.  The detections from the line search at
$1.0 < z < 1.7$ alone account for $\approx 40\%$ of the galaxy population at
$10 < \mathrm{SFR}[\Msun/\mathrm{yr}] < 30$, increasing to $>75\%$ at
$\mathrm{SFR} > 30 \Msun/\mathrm{yr}$.  Including the prior-based detections,
we find 60\% of the population at $\mathrm{SFR} \approx 20$~\Msun~yr$^{-1}$.
Similarly, at $2.0 < z < 3.1$, the detection fraction is highest in the most
highly star-forming bin.  Notably, however, with ASPECS-LP we probe molecular
gas in galaxies down to much lower SFRs as well.  The sources span over two
orders of magnitude in SFR, from $\approx 5$ to $>500$~\Msun~yr$^{-1}$.

The stellar masses of the ASPECS-LP detections in CO(2-1) and CO(3-2) are shown
in the left panels of \autoref{fig:hist-mstar-sfr}.  We detect molecular gas in
galaxies spanning over two orders of magnitude in stellar mass, down to
$\log M_{*}[\Msun] \sim 9.5$.  The completeness increases with stellar mass,
which is presumably a consequence of the fact that more massive galaxies
star-forming galaxies also have a larger gas fraction and higher SFR.  At
$M_{*} > 10^{10}$~\Msun, we are $\approx40\%$ complete at $1.0 < z < 1.7$,
while we are $\approx 50\%$ complete at $M_{*} > 10^{10.5}$~\Msun\ at
$2 < z < 3.1$.  The full distribution includes both star-forming and passive
galaxies, which would explain why we do not pick-up all galaxies at the highest
stellar masses.

\subsection{AGN fraction}
\label{sec:agn-fraction}
From the deepest X-ray data over the field we identify five AGN in the
ASPECS-LP line search sample (see \autoref{tab:derived-properties}).  Two of
these are detected in CO(2-1); namely, ASPECS-LP.3mm.05 and ASPECS-LP.3mm.15.
The remaining three X-ray AGN are ASPECS-LP.3mm.01, 3mm.09 and 3mm.12, detected
in CO(3-2).  The AGN fraction among the ASPECS-LP sources is thus $2/10 = 20\%$
at $1.0<z<1.7$ and $3/5 = 60\%$ at $2.0<z<3.1$ (note that including the
MUSE-prior sources decreases the AGN fraction).  If we consider the total
number of X-ray AGN over the field, we detect $2/6 = 30\%$ of the X-ray AGN at
$1.0<z<1.7$ and $3/6=50\%$ at $2.0<z<3.1$, without preselection.

The comoving number density of AGN increases out to $z\approx 2-3$
\citep{Hopkins2007}.  Using a volume limited sample out to $z\sim0.7$ based on
the Sloan Digital Sky Survey and \emph{Chandra}, \cite{Haggard2010} showed that
the AGN fraction increases with both stellar mass and redshift, from a few
percent at $M_{*} \sim 10^{10.7}$ \Msun, up to $20\%$ in their most massive bin
($M_{*} \sim 10^{11.8}$ \Msun).  Closer in redshift to the ASPECS-LP sample,
\cite{Wang2017} investigated the fraction of X-ray AGN in the GOODS fields and
found that among massive galaxies, $M_{*} > 10^{10.6} \Msun$, $5-15\%$ and
$15-50\%$ host an X-ray AGN at $0.5<z<1.5$ and $1.5<z<2.5$, respectively.  The
AGN fractions found in ASPECS-LP are broadly consistent with these ranges given
the limited numbers and considerable Poisson error.

Given the AGN fraction among the ASPECS-LP sources (20\% at $z\sim1.4$ and 60\%
at $z\sim2.6$), the question arises whether we detect the galaxies in CO
\emph{because} they are AGN (i.e., AGN-powered), or, whether we detect a
population of galaxies that hosts a larger fraction of AGN (e.g., because the
higher gas content fuels both the AGN and star-formation)?  The CO ladders in,
e.g., quasar host galaxies can be significantly excited, leading to an
increased luminosity in the high-$J$ CO transitions compared to star-forming
galaxies at lower excitation \citep[see, e.g.,][]{Carilli2013, Rosenberg2015}.
With the band 3 data we are sensitive to the lower-$J$ transitions, decreasing
the magnitude of such a bias towards AGN.  At the same time, the ASPECS-LP is
sensitive to the galaxies with the largest molecular gas reservoirs, which are
typically the galaxies with the highest stellar masses and/or SFRs.  As AGN are
more common in massive galaxies, it is natural to find a moderate fraction of
AGN in the sample, increasing with redshift.  Once the ASPECS-LP is complete
with the observations of the band 6 (1~mm) data, we can investigate the higher
$J$ CO transitions for these sources and possibly test whether the CO is
powered by AGN activity.

\subsection{Obscured and unobscured star formation rates}
\label{sec:star-formation-rates}

\begin{figure}[t]
  \includegraphics[width=\columnwidth]{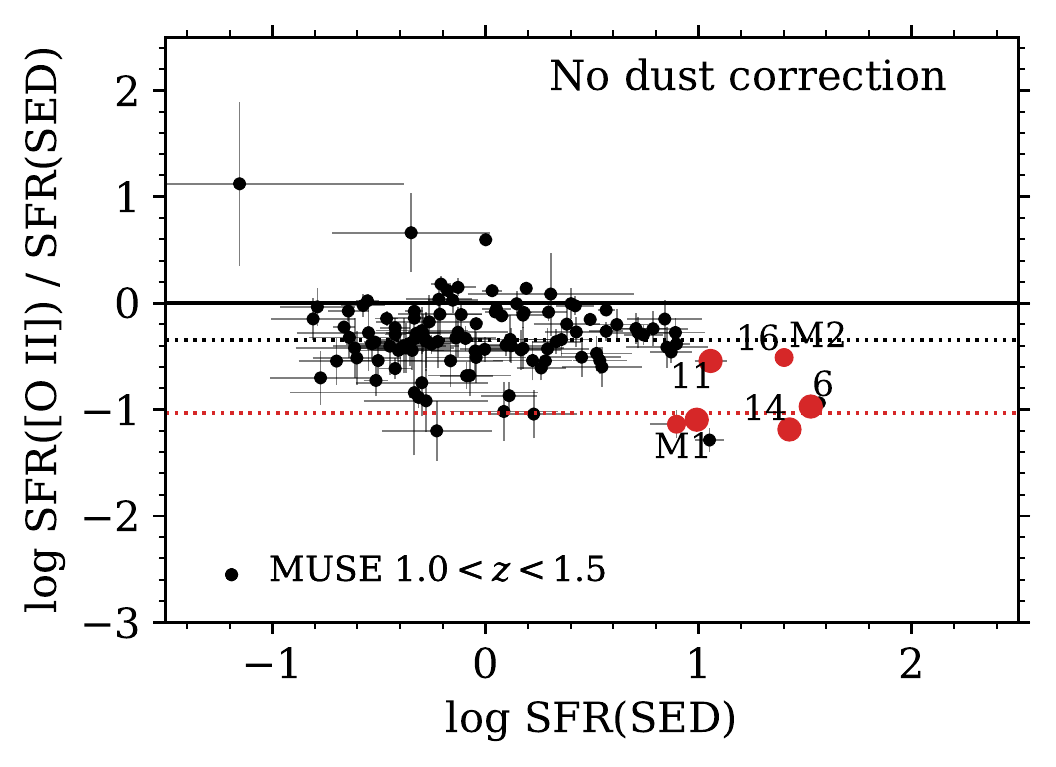}
  \caption{Total SFR from the SED fitting versus the ratio between SFR(\Oii)
    and SFR(SED) for the ASPECS-LP detected sources (red) and the MUSE
    $1.0<z<1.5$ reference sample (black).  This shows the ratio between the
    unobscured SFR(\OII) and total SFR.  The black and red dotted lines show
    the median ratio between SFR(\OII) and SFR(SED) for all galaxies and the
    ASPECS-LP sources only.  The median fraction of obscured/unobscured SFR is
    \fobsprior\ for all the ASPECS-LP sources. \label{fig:sfr-OII-1-1p5}}
\end{figure}
We investigate the fraction of dust-obscured star formation by comparing the
SFR derived from the \Oii\ emission line, without dust correction, with the
(independent) total SFR from modeling the UV-FIR SED with \textsc{magphys}.  We
show the ratio between the SFR(\OII) and the total SFR(SED) as a function of
the total SFR in \autoref{fig:sfr-OII-1-1p5}.  We use the observed (unobscured)
\OII\ luminosity, yielding a measurement of the fraction of unobscured SFR.
Immediately evident is the fact that more highly star-forming galaxies (which
are on average more massive) are more strongly obscured.  The median ratio
(bootstrapped errors) of obscured/unobscured SFR is \fobs\ for the ASPECS-LP
sources from the line search, which have a median mass of \fobsMmedian\
(cf. \fobsall\ for the complete sample of MUSE galaxies, with a median mass of
\fobsallMmedian).  Including the objects from the prior-based search does not
significantly affect this fraction (\fobsprior, at a median mass of
$M_{*} \sim \fobspriorMmedian$).

\subsection{Metallicities at $1.0 < z < 1.42$}
\label{sec:metallicities-at-1.0}
\begin{figure}[t]
  \includegraphics[width=\columnwidth]{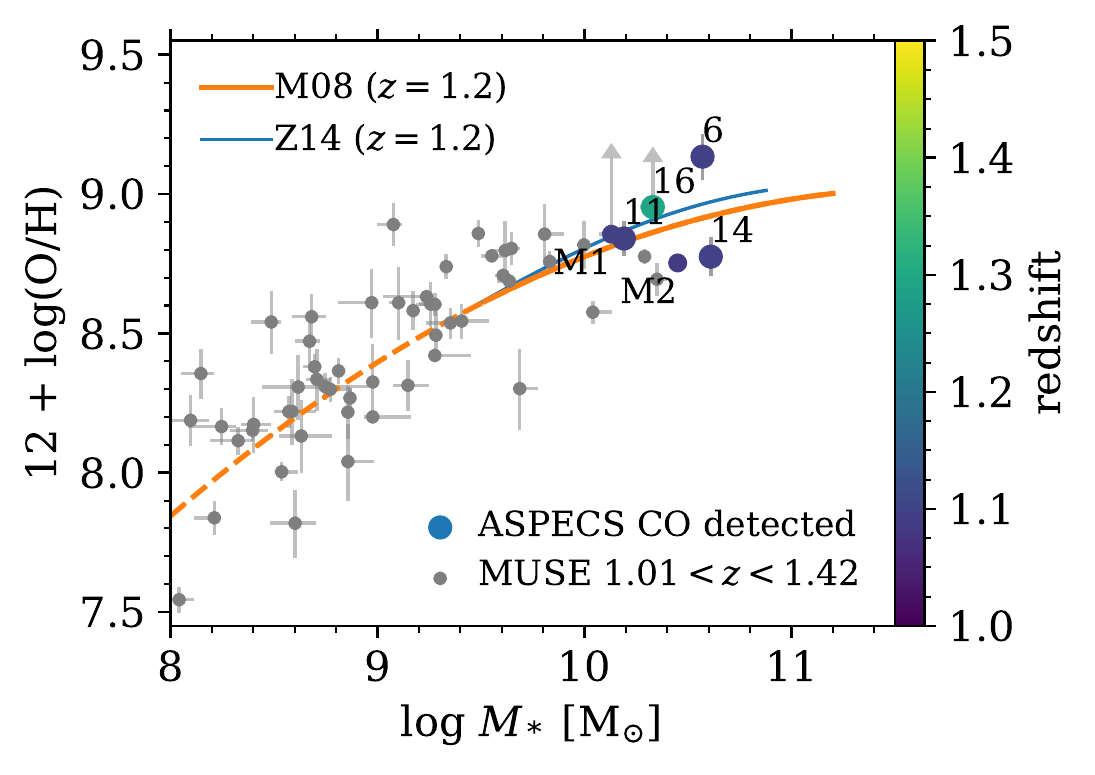}
  \caption{Stellar mass ($M_{*}$) - metallicity (\logOH) relation for the
    $1 < z < 1.5$ sub-sample.  We use the ratio of \Neiii\ and the
    \Oii-doublet, available at $z<1.42$, to derive the metallicity
    (\citealt{Maiolino2008}).  The solid lines show the mass-metallicity
    relations from \cite{Zahid2014} and \cite{Maiolino2008}, (converted to the
    same IMF and metallicity scale, \citealt{Kewley2008}), where the latter was
    interpolated to the average redshift of the sample (and extrapolated to
    lower masses, dashed line, for reference).  Overall, the ASPECS-LP galaxies
    are consistent with a (super-)solar metallicity.  \label{fig:MZ-1-1p5}}
\end{figure}
The molecular gas conversion factor is dependent on the metallicity, which is
therefore an important quantity to constrain.  Specifically, \aco\ can be
higher in galaxies with significantly sub-solar metallicities, where a large
fraction of the molecular gas may be CO faint, or lower in (luminous) starburst
galaxies, where CO emission originates in a more highly excited molecular
medium \citep[e.g.,][]{Bolatto2013}.

Given that the majority of the ASPECS-LP sources are reasonably massive,
$M_{*} \ge 10^{10}$~\Msun, their metallicities are likely to be (super-)solar,
based on the mass-metallicity relation \citep[e.g., ][]{Zahid2014}.

For the ASPECS-LP sources at $1.0 < z < 1.42$, the MUSE coverage includes
\Neiii, which can be used as a metallicity indicator
(\autoref{sec:metallicities}).  We infer a metallicity for ASPECS-LP.3mm.06,
3mm.11, 3mm.14 and ASPECS-LP-MP.3mm.02.  In addition, we can provide a lower
limit on the metallicity for ASPECS-LP.3mm.16 and ASPECS-LP-MP.3mm.01, based on
the upper limit on the flux of \NeIII.

In \autoref{fig:MZ-1-1p5}, we show the ASPECS-LP sources on the stellar mass -
gas-phase metallicity plane.  For reference, we show the mass-metallicity
relation from \cite{Maiolino2008} (that matches the \NeIII/\OII\ calibration)
and \cite{Zahid2014}, both converted to the same IMF and metallicity scale
\citep{Kewley2008}.  The AGN-free ASPECS-LP sources span about half a dex in
metallicity.  They are all metal-rich and consistent with a solar or
super-solar metallicity, in line with the expectations from the
mass-metallicity relation.

The (near-)solar metallicity of our targets supports our choice of $\aco = 3.6$
\Msun~(K~km~s$^{-1}$~pc$^2$)$^{-1}$, which was derived for $z\approx1.5$
star-forming galaxies \citep{Daddi2010} and is similar to the Galactic \aco\
\citep[cf.][]{Bolatto2013}.

\section{Discussion}
\label{sec:discussion}
\begin{figure}[t]
  \includegraphics[width=\columnwidth]{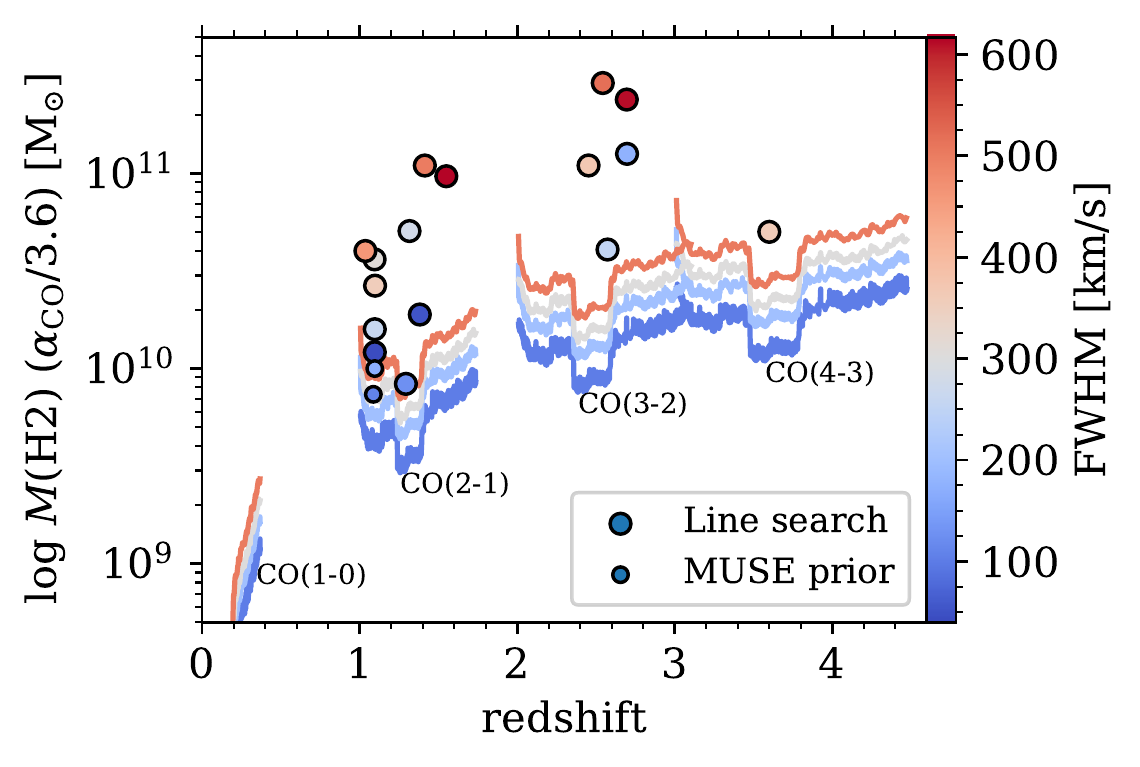}
  \caption{The $5\sigma$ molecular gas mass detection limit of ASPECS-LP as a
    function of redshift and CO line full-width at half-maximum (FWHM),
    assuming $\alpha_{\mathrm{CO}} = 3.6$ and \cite{Daddi2015} excitation (see
    \autoref{sec:molec-gas-prop}).  The sensitivity varies with redshift and
    increases with the square root of the decreasing line width at fixed
    luminosity, indicated by the color.  The points indicate the ASPECS-LP
    blind and prior-based sources, which are detected both in the deeper and
    shallower parts of the sensitivity curve.}
  \label{fig:flux-lim}
\end{figure}

\begin{figure*}[t]
  \includegraphics[width=\textwidth]{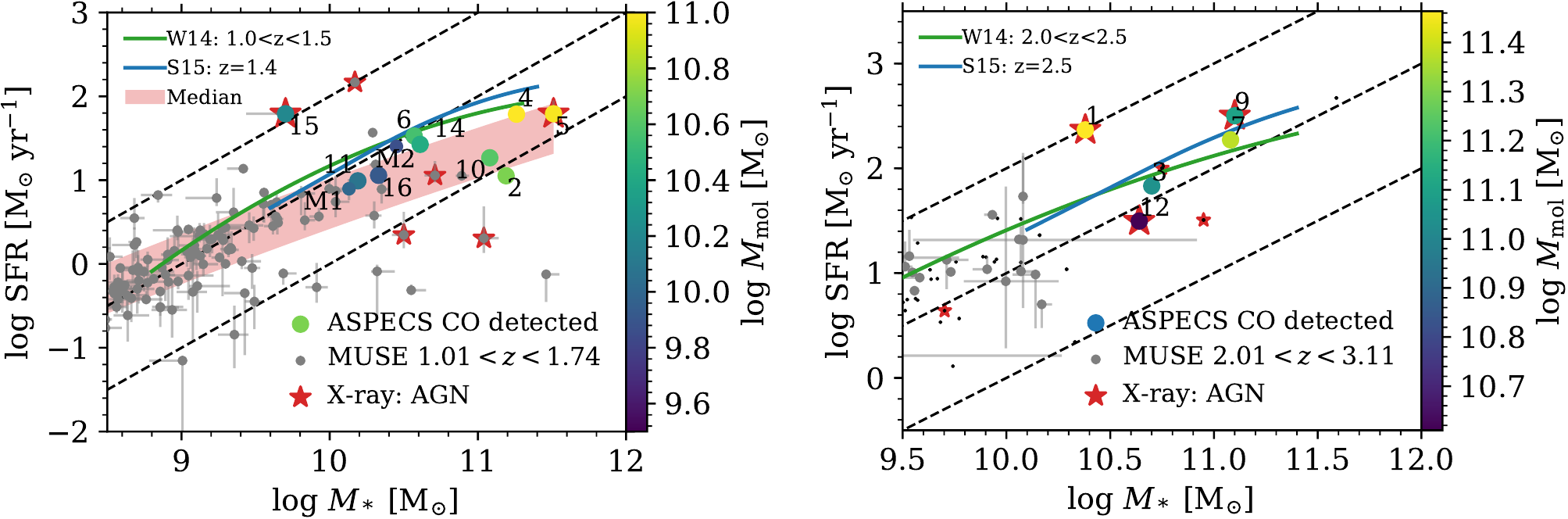}
  \caption{Stellar mass ($M_{*}$) versus SFR (from \textsc{Magphys}) for the
    CO(2-1) and CO(3-2) detected galaxies at $1.0 < z < 1.7$ (\textbf{left})
    and $2.0 < z < 3.1$ (\textbf{right}), respectively. The ASPECS-LP
    line-search and MUSE prior-based CO detections are represented by the
    larger and smaller circles respectively, colored by their molecular gas
    mass (\MH). The gray and black points show the MUSE and photometric
    reference sample of galaxies, respectively.  Red stars indicate X-ray
    sources identified as AGN from \cite{Luo2017}.  The green and blue solid
    curves denote the galaxy main sequence relationships from, respectively,
    \cite{Whitaker2014} and \cite{Schreiber2015}.  The red band shows
    $\pm0.3$~dex around a polynomial fit to the running median of all galaxies
    in the panel.  Lines of constant sSFR (0.1, 1 and 10 Gyr$^{-1}$) are shown
    black and dashed.  At $z\sim1.4$ ASPECS-LP detects molecular gas in
    galaxies that span a range of SFRs above, on and below the galaxy
    MS. \label{fig:msfr-1-3}}
\end{figure*}

\begin{figure*}[t]
  \includegraphics[width=\textwidth]{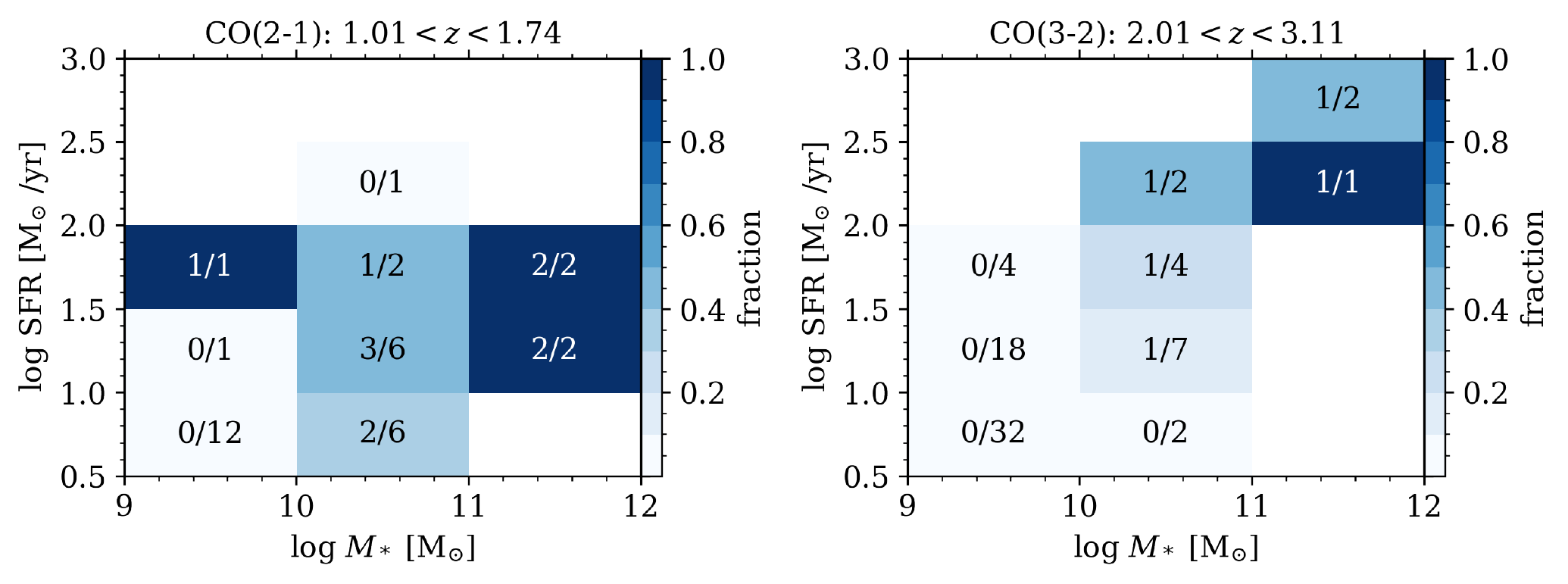}
  \caption{Fraction of sources detected by ASPECS-LP in $M_{*}$-SFR space at
    $1.01 < z < 1.74$ (\textbf{left}) and $2.01 < z < 3.11$ (\textbf{right}).
    This includes the detections from both the line search and the MUSE
    prior-based search.  We are most complete at the highest SFRs and stellar
    masses.  At a fixed stellar mass (SFR), the completeness fraction increases
    with SFR (stellar mass). \label{fig:det-frac}}
\end{figure*}

\begin{figure*}[t]
  \includegraphics[width=\textwidth]{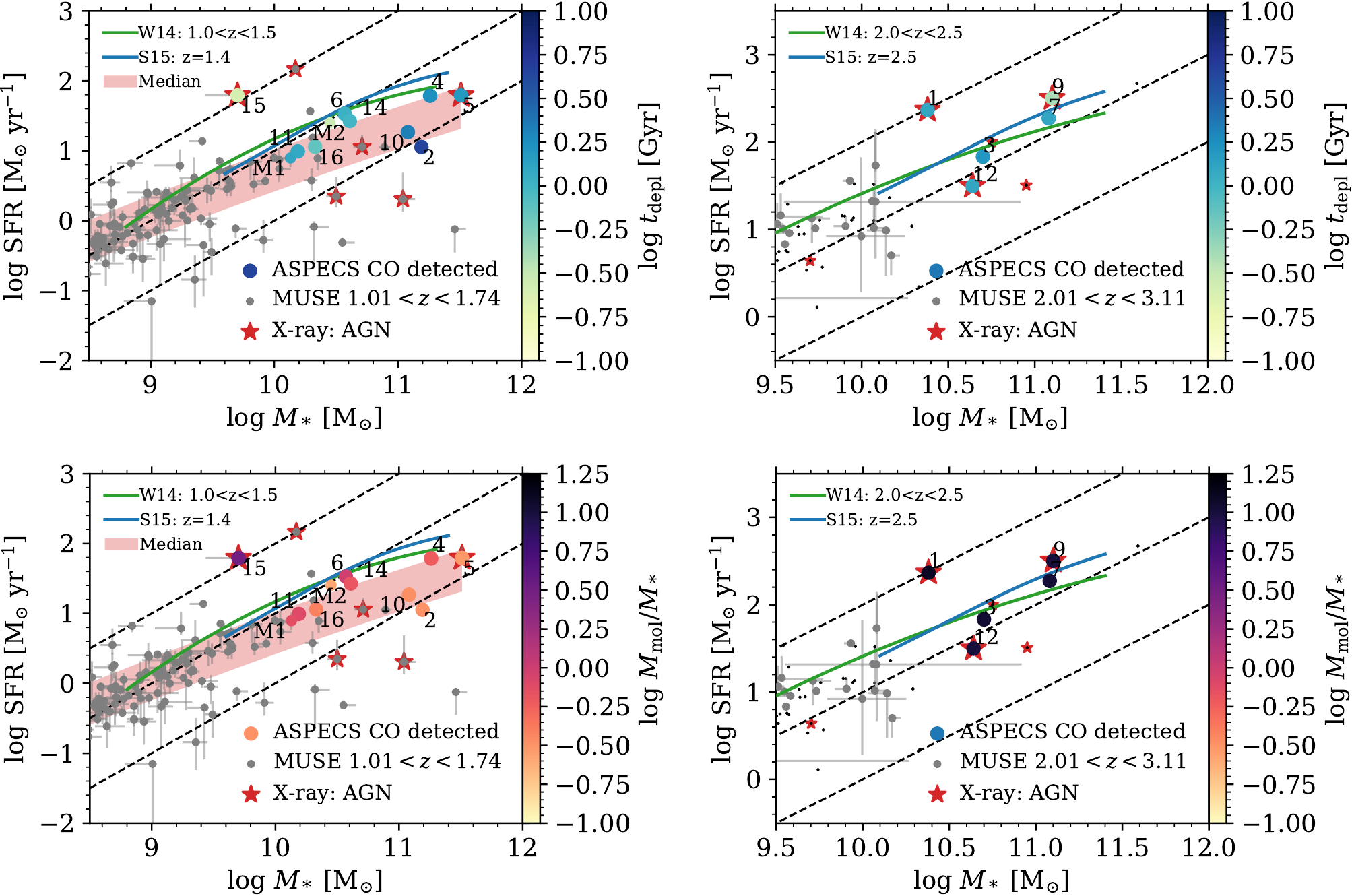}
  \caption{Stellar mass ($M_{*}$) versus SFR for the galaxies at $1 < z < 1.7$
    (\textbf{left}) and $2 < z < 3.1$ (\textbf{right}). The ASPECS-LP
    detections from the full and prior-based search are represented by the
    larger and smaller circles respectively. The gray and black points show the
    MUSE and photometric reference samples of galaxies, respectively.  Red
    stars indicate X-ray sources identified as AGN from \citep{Luo2017}.  The
    color in the different panels denotes the change in depletion time
    ($\tdepl = \MH/ \mathrm{SFR}$; \textbf{top}) and molecular-to-stellar
    mass ratio ($\MH/M_{*}$; \textbf{bottom}).  The green and blue solid curves
    denote the MS relationships from, respectively, \cite{Whitaker2014} and
    \cite{Schreiber2015}.  The red band shows $\pm0.3$~dex around a polynomial
    fit to the running median of all galaxies in the panel.  Lines of constant
    sSFR (0.1, 1 and 10 Gyr$^{-1}$) are shown black and dashed.  The gas
    fraction and depletion time vary systematically in galaxies across the main
    sequence.\label{fig:msfr-der-1-3} }
\end{figure*}

\subsection{Sensitivity limit to molecular gas reservoirs}
\label{sec:sens-limit-molec}
Being a flux limited survey, the limiting molecular gas mass of the ASPECS-LP,
$\MH(z)$, increases with redshift.  Based on the measured flux limit of the
survey, we can gain insight into what masses of gas we are sensitive to at
different redshifts.  The sensitivity of the ASPECS-LP Band 3 data itself is
presented and discussed in \cite{Gonzalez-LopezALP}\ (their Fig. 3): it is
relatively constant across the frequency range, being deepest in the center
where the different spectral tunings overlap.

Assuming a CO line full-width at half-maximum (FWHM) and an \aco\ and
excitation ladder as in \autoref{sec:molec-gas-prop}, we can convert the
root-mean-square noise level of ASPECS-LP in each channel to a sensitivity
limit on $\MH(z)$.  The result of this is shown in \autoref{fig:flux-lim}.
With increasing luminosity distance, ASPECS-LP is sensitive to more massive
reservoirs.  This is partially compensated by the fact that the first few
higher order transitions are generally more luminous at the typical excitation
conditions in star-forming galaxies.  The $\MH(z)$ function has a strong
dependence on the FWHM, as broader lines at the same total flux are harder to
detect (see also \citealt{Gonzalez-LopezALP}).  As the FWHM is related to the
dynamical mass of the system, and we are sensitive to more massive systems at
higher redshifts, these effects will conspire in further pushing up the
gas-mass limit to more massive reservoirs.

At $1.0 < z < 1.7$, the lowest gas mass we can detect at $5 \sigma$ (using the
above assumptions and a FWHM for CO(2-1) of 100 km~s$^{-1}$) is
$\MH \sim 10^{9.5}$ \Msun, with a median limiting gas mass over the entire
redshift range of $\MH \ge 10^{9.7}$ \Msun\ ($\MH \ge 10^{9.9}$ \Msun\ at FWHM
= 300 km~s$^{-1}$).  At $2.0 < z < 3.1$ the median sensitivity increases to
$\MH \gtrsim 10^{10.3}$ \solarmass, assuming a FWHM of 300 km~s$^{-1}$ for
CO(3-2).  In reality the assumptions made above can vary significantly for
individual galaxies, depending on the physical conditions of their ISM.

As cold molecular gas precedes star formation, the $\MH(z)$ selection function
of ASPECS-LP can, to first order, be viewed as a SFR$(z)$ selection function.
Since more massive star-forming galaxies have higher SFRs (albeit with
significant scatter), a weaker correlation with stellar mass may also be
expected.  These rough, limiting relations will provide useful context to
understand what galaxies we detect with ASPECS.

\subsection{Molecular gas across the galaxy main sequence}
\label{sec:molecular-gas-high}
We show the ASPECS-LP sources in the stellar mass - SFR plane at
$1.01 < z < 1.74$ and $2.01 < z < 3.11$ in \autoref{fig:msfr-1-3}.  On average,
star-forming galaxies with a higher stellar mass have a higher star formation
rate, with the overall star formation rate increasing with redshift for a given
mass, a relation usually denoted as the galaxy main sequence (MS).  We show the
MS relations from \citet[][W14]{Whitaker2014} and \citet[][S15]{Schreiber2015}
at the average redshift of the sample.  The typical intrinsic scatter in the MS
at the more massive end is around 0.3 dex or a factor 2 \citep{Speagle2014},
which we can use to discern whether galaxies are on, above or below the MS at a
given mass.

\subsubsection{Systematic offsets in the MS}
\label{sec:MS-systematics}
It is interesting to note that the average SFRs we derive with \textsc{Magphys}
are lower than what is predicted by the MS relationships from W14 and S15
(\autoref{fig:msfr-1-3}).  This offset is seen irrespective of including the
FIR photometry to the SED fitting of the ASPECS-LP sources.  This illustrates
the fact that different methods of deriving SFRs from (almost) the same data
can lead to somewhat different results (see, e.g., \citealt{Davies2016} for a
recent comparison).  Both W14 and S15 derive their SFRs by summing the
estimated UV and IR flux (UV+IR): W14 obtains the UV flux from integrating the
UV part of their best-fit FAST SED \citep{Kriek2009} and scales the
\emph{Spitzer}/MIPS 24~\micron\ flux tot a total IR luminosity using a single
template based on the \cite{Dale2002} models.  S15 instead uses (stacked)
\emph{Herschel}/PACS and SPIRE data for the IR luminosity, modeling these with
\cite{Chary2001} templates.

Recently, \cite{Leja2018} remodeled the UV--24\micron\ photometry for all
galaxies from 3D-HST survey (which were used in deriving the W14 result) using
the Bayesian SED fitting code \textsc{Prospector-}$\alpha$ \citep{Leja2017}.
While \textsc{Prospector}-$\alpha$ also models the broadband SED in a Bayesian
framework and shares several similarities with \textsc{Magphys}, such as the
energy-balance assumption, it is a completely independent code with its own
unique features (e.g., the inclusion of emission lines, different stellar
models and non-parametric star formation histories).  Interestingly, the SFRs
derived by \cite{Leja2018} are $\sim0.1 - 1$~dex lower than those derived from
UV+IR, because of the contribution of old stars to the overall energy output
that is neglected in SFR(UV+IR).

While the exact nature of this offset remains to be determined, solving the
systematic calibrations between different SFR indicators (or a rederivation of
the MS relationship) is beyond the scope of this paper.  In the following we
show $\pm0.3$~dex scatter around a second order polynomial fit to the rolling
median of the SFRs all the galaxies (without any color selection) as a
reference in the lower redshift bin.  At the massive end where our ASPECS-LP
sources lie, we indeed find that this curve lies somewhat below the W14 and S15
relationships.  In the higher redshift bin the situation is less clear (given
the limited number of sources) and we keep the literature references.  With
this description of the median SFR at a given stellar mass in hand, we are in
the position to compare the SFRs of the individual ASPECS-LP sources to the
population average SFR.

\subsubsection{Normal galaxies at $1.01 < z < 1.74$}
\label{sec:normal-galaxies-at-1-z-2}
The ASPECS-LP sources at $1.01 < z < 1.74$ are shown in the ($M_{*}$,
SFR)-plane in left panel of \autoref{fig:msfr-1-3}.  For comparison we show all
sources in this redshift range with a secure spectroscopic redshift, as MUSE is
mostly complete for massive, star-forming galaxies in the regime of the
ASPECS-LP detections at these redshifts (see \autoref{fig:hist-mstar-sfr}).

At the depth of the ASPECS-LP, we are sensitive enough to probe molecular gas
reservoirs in a variety of galaxies that lie on and even below the MS at
$z\sim1.4$.  Most of the ASPECS-LP galaxies detected in this redshift range lie
on the main sequence, spanning a mass range of $\sim2$ decades at the massive
end.  These galaxies belong to the population of normal star-forming galaxies
at these redshifts.

As expected, with the primary sample alone we detect essentially all massive
galaxies that lie above the main sequence, for $M_{*} > 10^{9.5}$ \Msun.  The
lowest mass galaxy we detect is ASPECS-LP.3mm.15, which is elevated
significantly above the MS and is also an X-ray classified AGN.  One galaxy,
with the highest SFR of all, is a notable outlier for not being detected:
MUSE\#872 ($M_{*} = 10^{10.2} $ \Msun, SFR $\sim 150$ \Msun~yr$^{-1}$).  From
the prior-based search we find that no molecular gas emission is seen in this
source at lower levels either.  While the non-detection of this object is very
interesting, we caution that this source is also a broad-line AGN in MUSE and
it is possible that its SFR is overestimated.

Notably, we also detect a number of galaxies that lie significantly below the
main sequence (e.g., ASPECS-LP.3mm.02), meaning they have SFRs well below the
population average.  Despite their low SFR, these sources host a significant
gas reservoir and have a gas fraction that is in some cases similar to MS
galaxies at their stellar mass.  The detection of a significant molecular gas
reservoir in these sources is interesting, as these sources would typically not
be selected in targeted observations for molecular gas.

Overall, we detect the majority of the galaxies on the massive end of the MS at
$1.0 < z < 1.7$ in CO.  We show the detection rate in bins of stellar mass and
SFR in the left panel of \autoref{fig:det-frac}.  At a SFR $> 10 $
\Msun~yr$^{-1}$, we detect $>60\%$ of galaxies at all masses at these
redshifts.  If we focus on galaxies with $M_{*} > 10^{10} \Msun$, we are
$>60\%$ complete down to log~SFR[\Msun~yr$^{-1}$]~$>0.5$, where we encompass
all MS galaxies.

\subsubsection{Massive galaxies at $2.01 < z < 3.11$}
\label{sec:massive-galaxies-at-2-z-3}
At $2.01 < z < 3.11$, we are sensitive to CO(3-2) emission from massive gas
reservoirs.  We plot the galaxies detected in CO(3-2) on the main sequence in
the right panel of \autoref{fig:msfr-1-3}.  For completeness, we have added
ASPECS-LP.3mm.12 to the figure as well, but caution that the photometry is
blended with a lower redshift foreground source.  As the number of
spectroscopic redshifts from MUSE is more limited in this regime, we also
include galaxies from our extended photometric redshift catalog as small black
dots (indicating AGN with red stars).

The detections from ASPECS-LP make up most of the massive and highly
star-forming galaxies at these redshifts.  Based on their CO flux, the sources
all have a molecular gas mass of $\geq 10^{10.5}$ \Msun\ and correspondingly
high molecular gas fractions $\MH/M_{*} \ge 1.0$.  Their SFRs differ by over an
order of magnitude.  ASPECS-LP.3mm.07 and 09 are both at $z\approx2.697$ and
lie on the MS with SFRs between $150-350$ \Msun~yr$^{-1}$.  In contrast,
ASPECS-LP.3mm.03 has a lower SFR of $<100$\Msun~yr$^{-1}$.  ASPECS-LP.3mm.01
has a very high gas fraction and SFR for its stellar mass and is also detected
as an X-ray AGN.

We show the quantitative detection fraction for CO(3-2) at these redshifts in
the right panel of \autoref{fig:det-frac}.  Note that, as the area of the HUDF
and the ASPECS-LP is small, there are relatively few massive galaxies in the
field at these redshifts.

\subsection{Evolution of molecular gas content in galaxies}
\label{sec:evol-molec-gas}
We now provide a brief discussion of the evolution of the molecular gas
properties (and the individual outliers) in the full ASPECS-LP sample of 18
sources, including the muse prior based sources, in the context of the MUSE
derived properties.  A more detailed discussion of these results will be
provided in \cite{AravenaALP}.

From systematic surveys of the galaxy population at $z\approx0$, we know that
the molecular gas properties of galaxies vary across the main sequence
\citep[e.g.,][]{Saintonge2016, Saintonge2017}.  The same trends are unveiled in
the ASPECS-LP sample out to $z\approx3$.  To reveal these trends more clearly,
we show the main sequence plot colored by the depletion time
($t_{\rm depl} = \MH/\mathrm{SFR}$) and gas fraction (indicated by $\MH/M_{*}$)
in \autoref{fig:msfr-der-1-3}.  The molecular gas mass and depletion time of
the ASPECS-LP sources vary systematically across the MS.  On average, galaxies
above the MS have higher gas fractions and shorter depletion times than
galaxies on the MS, while the contrary is true for galaxies below the MS
(longer depletion times, smaller gas fractions).

At $z\sim1.4$, the sources span about an order of magnitude in depletion time,
from $0.3 - 5$~Gyr, with a median depletion time of \medTdeplZone~Gyr.  This
comparable to the average depletion times found in $z=1-3$ star-forming
galaxies \citep[e.g.,][]{Daddi2010, Tacconi2013}.  ASPECS-LP.3mm.02, which
appears to harbor a substantial gas reservoir while its SFR puts it
significantly below the main sequence, has a correspondingly long depletion
time of several Gyr.  Although the numbers are more limited at higher
redshifts, we see a similar variety in depletion times at $z\sim2.6$, with a
median depletion time of \medTdeplZtwo~Gyr.  For galaxies of similar masses we
do not find a strong evolution in the depletion time between the $z\sim1.4$ and
$z\sim2.6$ bins.

The evolution of the gas fraction across the MS is clearly seen for the sources
at $z\sim1.4$.  The lowest gas-mass fractions we find are of the order of 30\%,
while the galaxies with the highest gas fractions have about equal mass in
stars and in molecular gas, with a median of $\MH/M_{*} \medFgasZone$.  These
are comparable to the gas fractions found at similar redshifts
\citep{Daddi2010, Tacconi2013}.  The gas fractions at $z\sim2.6$ are
substantially higher than they are at lower redshift.  ASPECS-LP.3mm.09 and 12
have substantial gas fractions close to unity, while both ASPECS-LP.3mm.03 and
07, have a molecular gas mass about $\times2$ their mass in stars (median
$\MH/M_{*} \medFgasZtwo$).  ASPECS-LP.3mm.1, 3mm.13 and 3mm.15 are outliers in
this picture, with a substantially higher gas fraction than the other sources.
Both 3mm.01 and 3mm.15 are also starbursts with a high inferred SFR and show an
X-ray detected AGN.  This high SFR is consistent with the high gas fraction and
a picture in which the large gas reservoir fuels a strong starburst, while some
gas powers the AGN simultaneously.  As may be expected given the flux-limited
nature of the observations, the highest redshift source, ASPECS-LP.3mm.13, also
has a substantial gas fraction ($\MH/M_{*} = \medFgasZthree$).  As a whole,
\autoref{fig:msfr-der-1-3} reveals the strength of the ASPECS-LP probing the
molecular gas across cosmic time without preselection.

\section{Summary} \label{sec:conclusions}

In this paper we use two spectroscopic integral-field observations of the
Hubble Ultra Deep Field, ALMA in the millimeter, and MUSE in the optical
regime, to further our understanding of the properties of the galaxy population
at the peak of cosmic star formation ($1 < z < 3$).  We start with the line
emitters identified from the ASPECS-LP Band 3 (3~mm) data without any
preselection \citep{Gonzalez-LopezALP}.  By using the MUSE data, as well as the
deep multi-wavelength data that is available for the HUDF, we find that all
ALMA-selected sources are associated with a counterpart in the optical/near-IR
imaging.  The spectroscopic information from MUSE enables us to associate all
ALMA line emitters with emission coming from rotational transitions of carbon
monoxide (CO) that result in unique redshift identifications: We identify 10
line emitters as CO(2-1) at $1 < z < 2$, five as CO(3-2) at $2 < z < 3$ and one
as CO(4-3) at $z=3.6$.  The line search done using the ALMA data is
conservative, to avoid contamination by spurious sources in the very large 3~mm
data cube \citep{Gonzalez-LopezALP}.  We therefore also use the MUSE data as a
positional and redshift prior to push the detection limit of the ALMA data to
greater depth and identify additional CO emitters at $z < 2.9$, increasing the
total number of ALMA line detections in the field to 18.

We present MUSE spectra of all CO-selected galaxies, and use the diagnostic
emission lines covered by MUSE to constrain the physical properties of the ALMA
line emitters.  In particular, for galaxies with coverage of \Oiialt/\Neiii\ at
$z\le1.5$ in the MUSE data, we infer metallicities consistent with being
(super-)solar, which motivates our choice of a Galactic conversion factor to
transform CO luminosities to molecular (H$_2$) gas masses for these galaxies in
this series of ASPECS-LP papers \citep{DecarliALP, Gonzalez-LopezALP,
  AravenaALP, PoppingALP}.  We also compare the unobscured \Oiialt-derived star
formation rates of the galaxies to the total SFR derived from their spectral
energy distributions with \textsc{Magphys} and confirm that a number of them
have high extinction in the rest-frame UV/optical regime.

Using the very deep \emph{Chandra} imaging available for the HUDF, we determine
an X-ray AGN fraction of 20\% and 60\% among the CO(2-1) and CO(3-2) emitters
at $z\sim1.4$ and $z\sim2.6$, respectively, suggesting that we do not
preferentially detect AGN at $z<2$.  A future analysis of the band 6 data from
the ASPECS-LP will reveal if those sources hosting an AGN show higher CO
excitation compared to those that do not.

We use the exquisite multi-wavelength data available for the HUDF to derive
basic physical parameters (such as stellar masses and star formation rates) for
all galaxies in the HUDF.  We recover the main sequence of galaxies and show
that most of our CO detections are located towards higher stellar masses and
star formation rates, consistent with expectations from earlier studies.
However, being a CO-flux limited survey, besides galaxies on or above the main
sequence our ALMA data also reveal molecular gas reservoirs in galaxies below
the main sequence at $z\sim1.4$, down to star formation rates of $\approx 5$
\Msun~yr$^{-1}$ and stellar masses of $M_{*} \approx 10^{10}$ \Msun.  At higher
redshift, we detect massive and highly star-forming galaxies in molecular gas
emission on and above the MS.  With our ALMA spectral scan, for stellar masses
$M_{*} \ge 10^{10} (10^{10.5})$~\Msun, we detect about $40\%$ (50\%) of all
galaxies in the HUDF at $1 < z < 2$ ($2 < z < 3$).  The ASPECS-LP galaxies span
a wide range of gas fractions and depletion times, which vary with their
location above, on and below the galaxy main-sequence.

The cross-matching of the integral-field spectroscopy from ALMA and MUSE has
enabled us to perform an unparalleled study of the galaxy population at the
peak of galaxy formation in the HUDF.  Given the large range of redshifts
covered by the ALMA spectral lines, key diagnostic lines in the UV/optical are
only covered by the MUSE observations in specific redshift ranges.  The launch
of the {\em James Webb Space Telescope} will greatly expand the coverage of
spectral lines that will help to further constrain the physical properties of
ALMA-detected galaxies in the HUDF.

\acknowledgments

We are grateful to the referee for providing a constructive report.
L.A.B. wants to thank Madusha L.P. Gunawardhana for her help with
\textsc{platefit}.  Based on observations collected at the European Southern
Observatory under ESO programme(s): 094.A-2089(B), 095.A-0010(A),
096.A-0045(A), and 096.A-0045(B).  This paper makes use of the following ALMA
data: ADS/JAO.ALMA\#2016.1.00324.L.  ALMA is a partnership of ESO (representing
its member states), NSF (USA) and NINS (Japan), together with NRC (Canada), NSC
and ASIAA (Taiwan), and KASI (Republic of Korea), in cooperation with the
Republic of Chile. The Joint ALMA Observatory is operated by ESO, AUI/NRAO and
NAOJ. The National Radio Astronomy Observatory is a facility of the National
Science Foundation operated under cooperative agreement by Associated
Universities, Inc.

``Este trabajo cont\'o con el apoyo de CONICYT + Programa de Astronom\'ia+
Fondo CHINA-CONICYT'' J.G-L. acknowledges partial support from ALMA-CONICYT
project 31160033. F.E.B. acknowledges support from CONICYT grant Basal
AFB-170002 (FEB), and the Ministry of Economy, Development, and Tourism's
Millennium Science Initiative through grant IC120009, awarded to The Millennium
Institute of Astrophysics, MAS (FEB).  J.B. acknowledges support by Funda{\c
  c}{\~a}o para a Ci{\^e}ncia e a Tecnologia (FCT) through national funds
(UID/FIS/04434/2013) and Investigador FCT contract
IF/01654/2014/CP1215/CT0003., and by FEDER through COMPETE2020
(POCI-01-0145-FEDER-007672).  T.D-S. acknowledges support from ALMA-CONYCIT
project 31130005 and FONDECYT project 1151239.  J.H. acknowledges support of
the VIDI research programme with project number 639.042.611, which is (partly)
financed by the Netherlands Organization for Scientific Research (NWO)
D.R. acknowledges support from the National Science Foundation under grant
number AST-1614213.  I.R.S. acknowledges support from the ERC Advanced Grant
DUSTYGAL (321334) and STFC (ST/P000541/1)

Work on \textsc{Gnuastro} has been funded by the Japanese MEXT scholarship and
its Grant-in-Aid for Scientific Research (21244012, 24253003), the ERC advanced
grant 339659-MUSICOS, European Union’s Horizon 2020 research and innovation
programme under Marie Sklodowska-Curie grant agreement No 721463 to the SUNDIAL
ITN, and from the Spanish MINECO under grant number AYA2016-76219-P.

\vspace{5mm}

\facilities{ALMA, VLT:Yepun}

\software{\textsc{Topcat} \citep{Taylor2005}, \textsc{Gnuastro}
  \citep{Akhlaghi2015}, \textsc{IPython} \citep{Perez2007}, \textsc{numpy}
  \citep{VanDerWalt2011}, \textsc{Matplotlib} \citep{Hunter2007},
  \textsc{Astropy} \citep{Robitaille2013, TheAstropyCollaboration2018}.}

\appendix

\section{Source description and redshift identifications}
\label{sec:redsh-conf}
\begin{figure*}[t]
  \includegraphics[width=\textwidth]{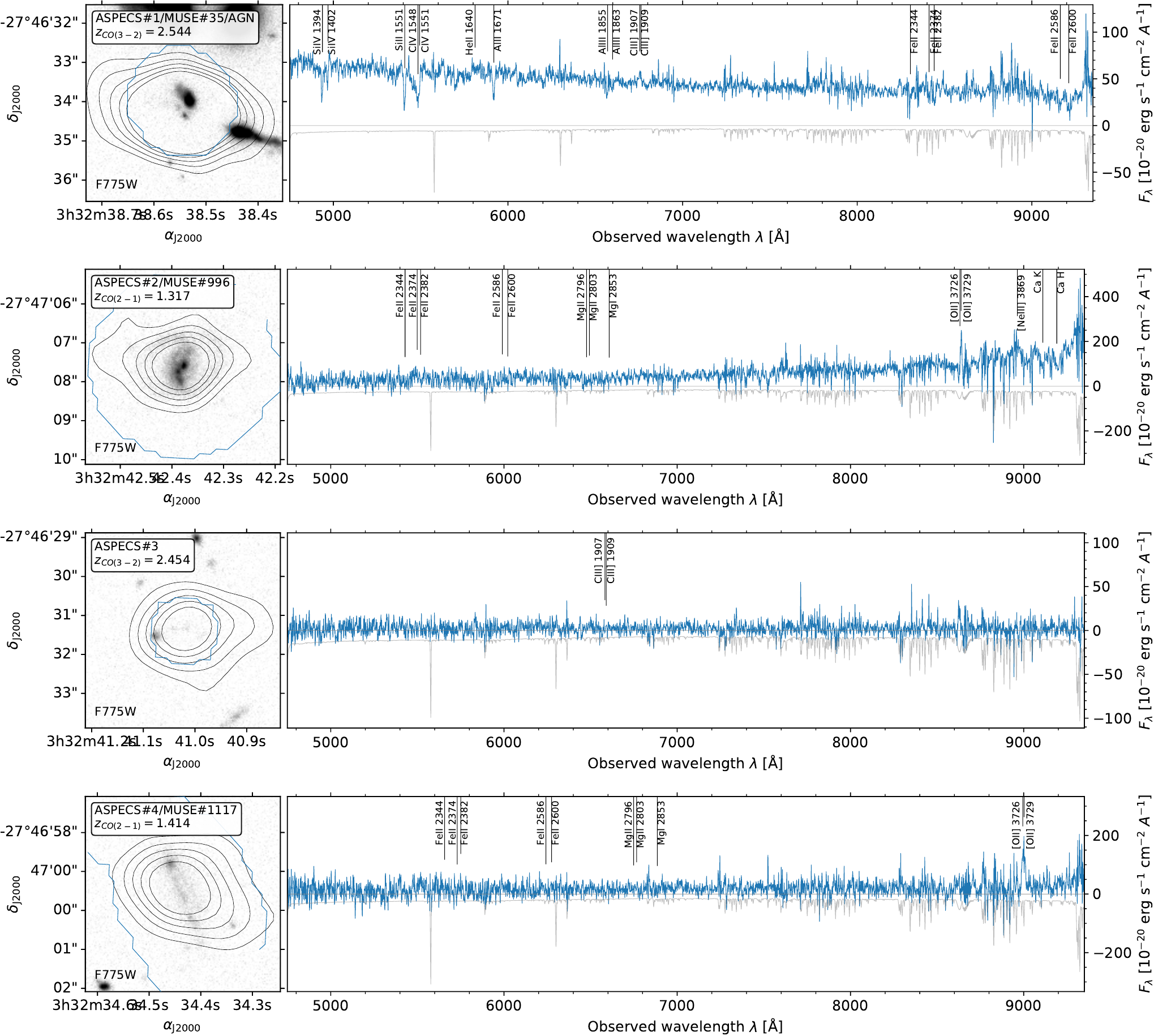}
  \caption{MUSE spectroscopy of the ASPECS-LP sources.  \textbf{Left:}
    \emph{HST}/F775W cutout.  The colored contour(s) mark the region of the
    spectral extraction(s), defined by convolving the MUSE PSF with the
    \emph{HST} segmentation map (see \citealt{Inami2017} for details), or a
    fixed 0.8\arcsec\ aperture (in the case of a new extraction; \#3, \#7, \#9,
    \#13).  The black contours indicate the CO emission from
    $\pm[3, .., 11]\sigma$ in steps of $2\sigma$.  \textbf{Right:} MUSE
    spectrum (1.5~\AA\ Gaussian smoothing) extracted over the marked region.
    The 1$\sigma$ uncertainty on the spectrum is shown by the gray line in the
    direction of negative flux (for clarity), and is largest around skylines.
    The (expected) positions of different spectral features are annotated; this
    does not indicate that the feature is also detected.  Spectra and lines of
    nearby or blended sources are shown in red. \label{fig:MUSE-spectra-1}}
\end{figure*}

\begin{figure*}[t]
  \includegraphics[width=\textwidth]{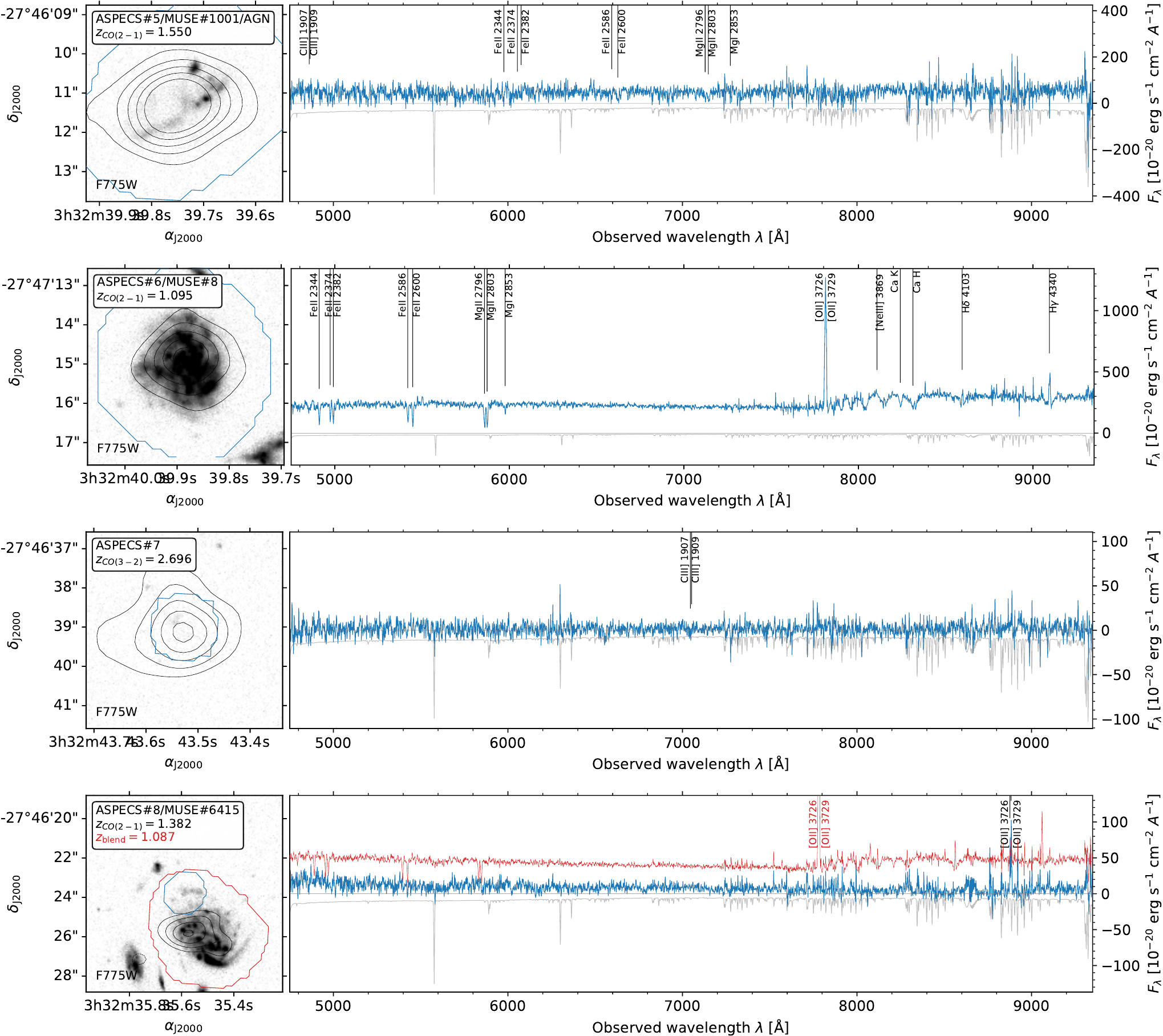}
  \caption{Continuation of \autoref{fig:MUSE-spectra-1}.  We show spectrum of
    the complete system of spiral galaxies at ASPECS-LP.3mm.08 in red (scaled
    down by a factor of 10), to make the faint \Oii\ line matching the CO(2-1)
    redshift visible.  The spectrum for ASPECS-LP.3mm.08 itself is shown in
    blue (extracted only over a part of the spiral arm).  Note the foreground
    source is independently detected in CO(2-1) as ASPECS-LP-MP.3mm.02 and the
    fully annotated spectrum for this source is show in
    \autoref{fig:MUSE-spectra-3}.\label{fig:MUSE-spectra-2}}
\end{figure*}

\begin{figure*}[t]
  \includegraphics[width=\textwidth]{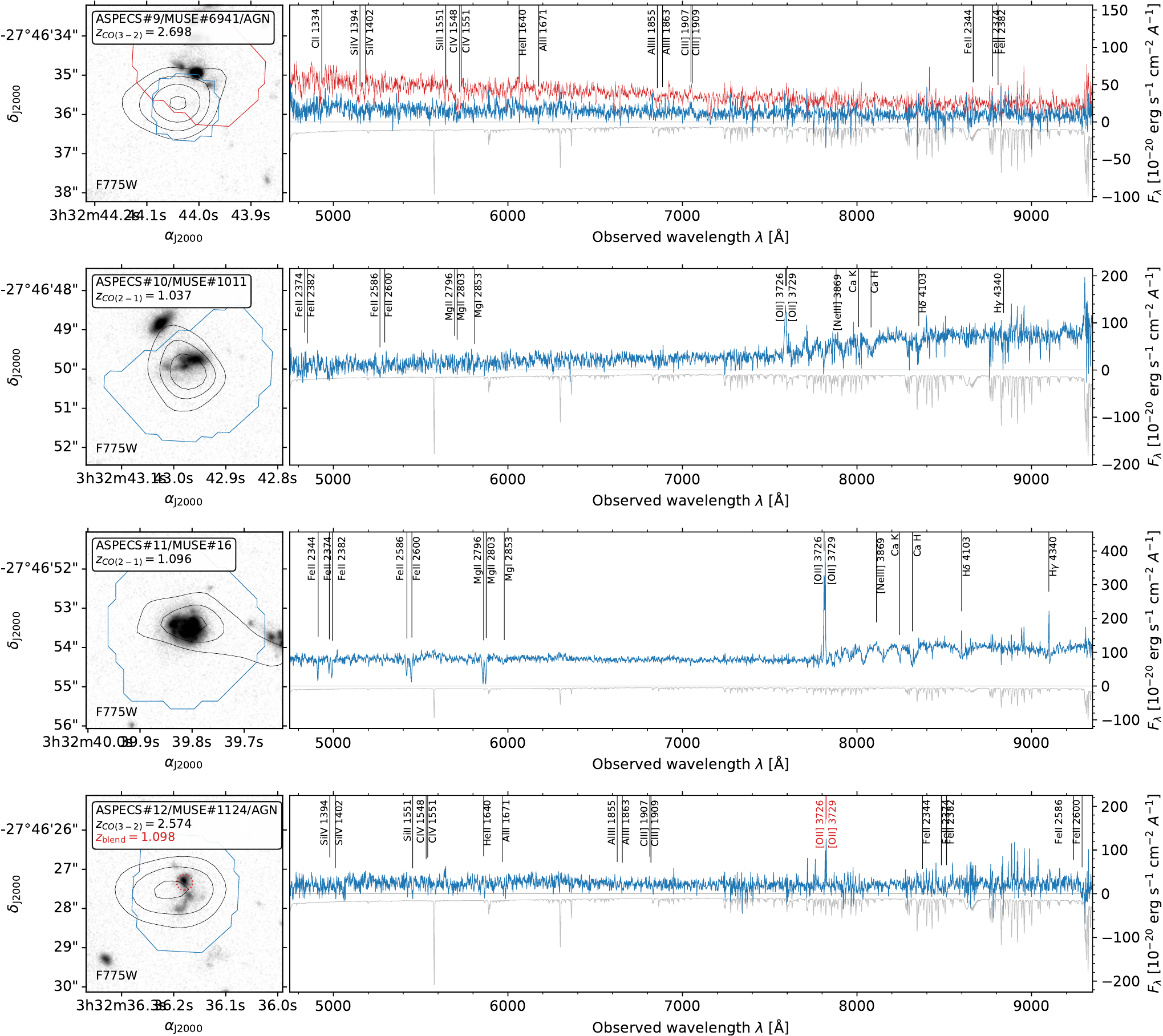}
  \caption{Continuation of \autoref{fig:MUSE-spectra-1}.  For ASPECS-LP.3mm.09
    the UV absorption features matching the CO(3-2) redshift are seen in the
    source to the north (red spectrum). ASPECS-LP.3mm.12 is blended with a
    foreground \OII-emitter (see
    \autoref{fig:ASPECS12}).\label{fig:MUSE-spectra-3}}
\end{figure*}

\begin{figure*}[t]
  \includegraphics[width=\textwidth]{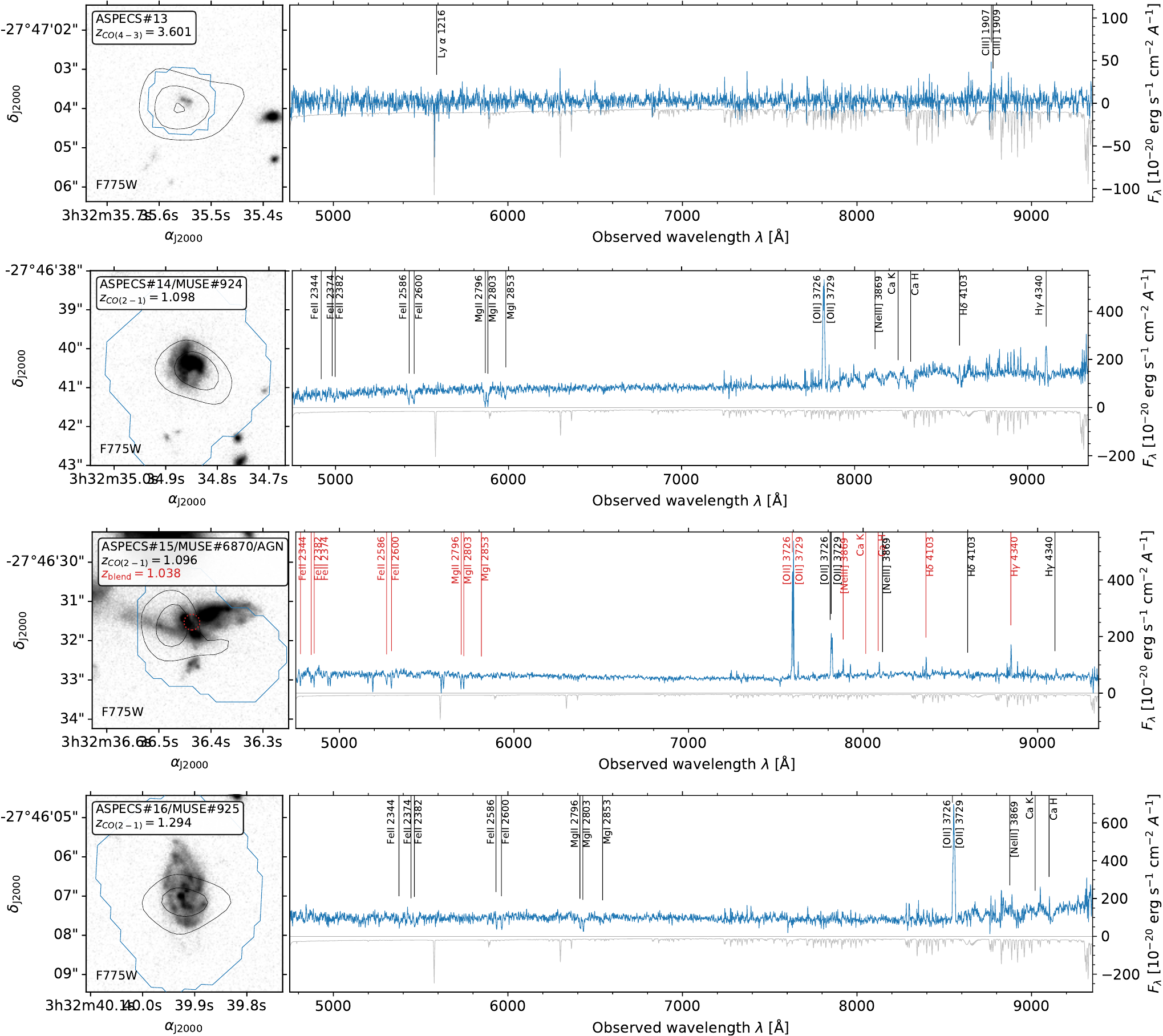}
  \caption{Continuation of \autoref{fig:MUSE-spectra-1}.  The spectrum of
    ASPECS-LP.3mm.15 is severely blended.  We have highlighted the strongest
    blended features (from a foreground source at $z=1.038$) in
    red.\label{fig:MUSE-spectra-4}}
\end{figure*}

\begin{figure*}[t]
  \includegraphics[width=\textwidth]{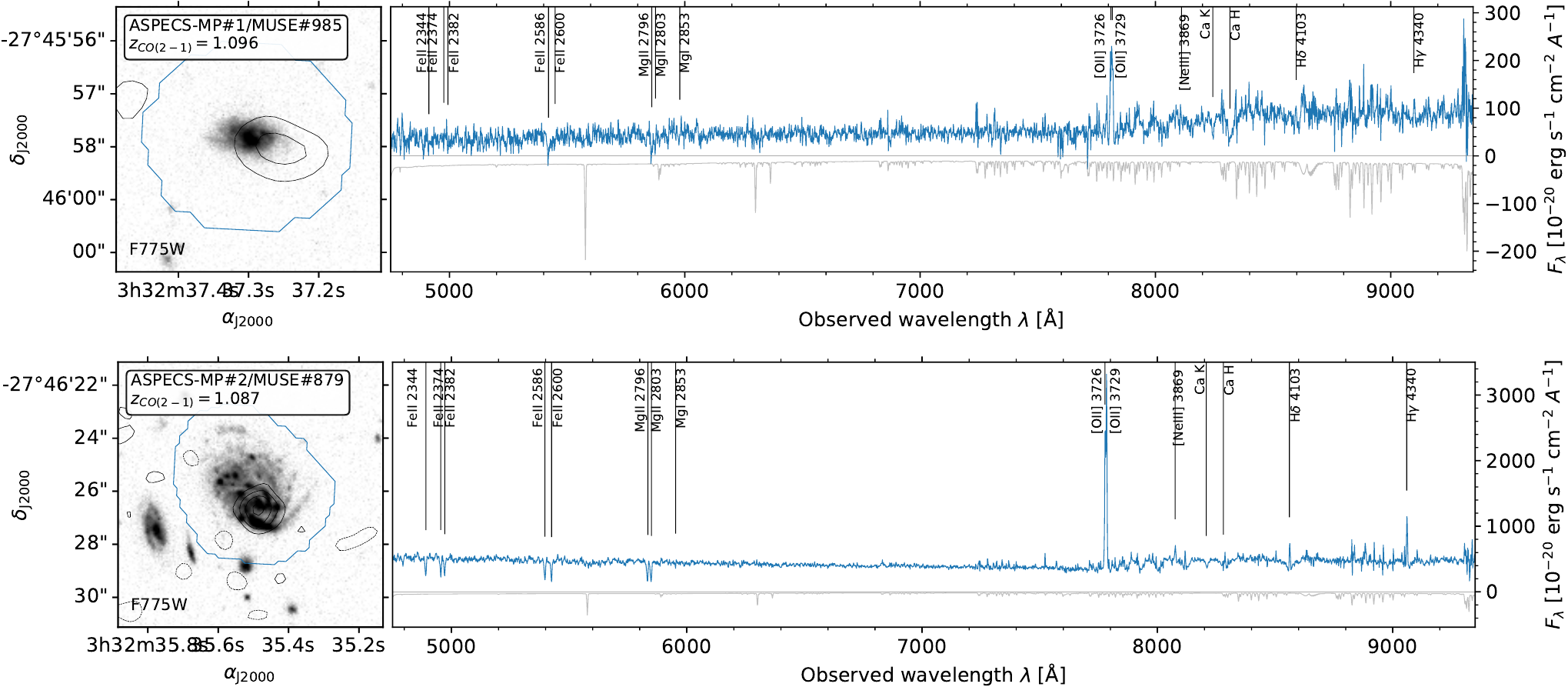}
  \caption{Continuation of \autoref{fig:MUSE-spectra-1}, showing the MUSE-prior
    based sources.\label{fig:MUSE-spectra-5}}
\end{figure*}

\begin{figure}[t]
  \includegraphics[width=0.5\columnwidth]{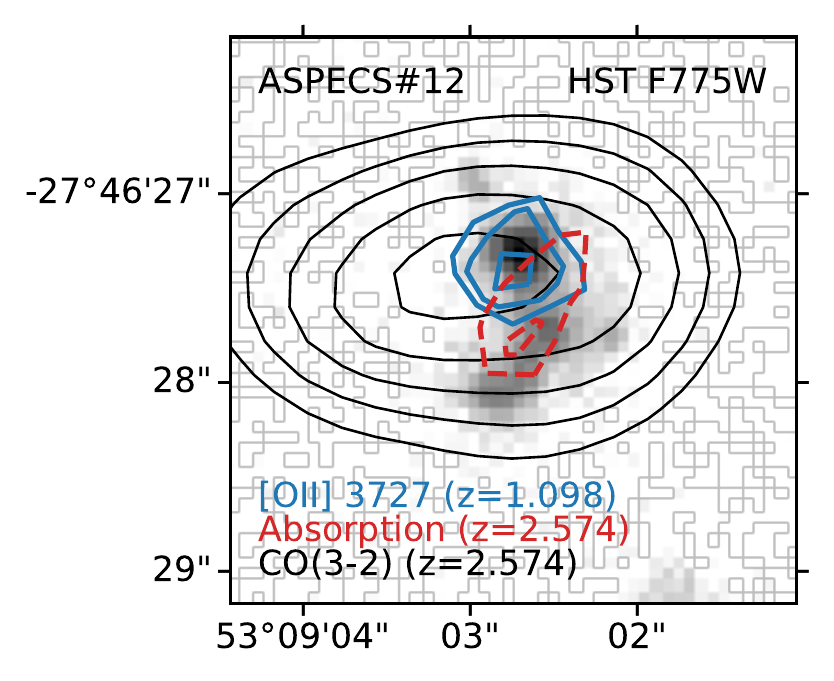}
  \caption{\emph{HST}/F775W cutout of ASPECS-LP.3mm.12, with CO(3-2) emission
    at $z=2.5738$.  The MUSE spectrum of this source reveals two redshifts.
    There is \OII\ emission at z=1.098, spatially consistent with a foreground
    galaxy to the north.  While the continuum is faint, cross-correlating the
    spectrum with an absorption line template reveals a peak at $z=2.5738$ with
    S/N~$> 4$.  Subsequent stacking of the UV absorption lines (\Cii, \Siiv,
    \Siii, \Civ, \Alii, \Aliii) reveals that the absorption is
    co-spatial with the background galaxy and consistent with the redshift of
    CO(3-2).  All contours start at $\pm 3 \sigma$, increasing by $1\sigma$
    (where solid and dashed indicate emission and absorption
    respectively). \label{fig:ASPECS12}}
\end{figure}

ASPECS-LP.3mm.01: CO(3-2) at $z=2.543$.  The brightest CO line emitter in the
field.  It is a \emph{Chandra}/X-ray detected AGN \citep[][\#718]{Luo2017} and
was already found in the line search at 3 mm and 1 mm in CO(3-2), CO(7-6) and
CO(8-7) and continuum in the ASPECS-Pilot (\citealt{Walter2016, Decarli2016},
3mm.1, 1mm.1, 1mm.2; \citealt{Aravena2016}, C1) as well as at 1 mm and 5 cm
continuum \citep[][UDF3]{Dunlop2017, Rujopakarn2016}.  The MUSE spectrum
(MUSE\#35) reveals a high S/N continuum with a wealth of UV absorption features
and \Ciii\ emission, confirming the redshift (\autoref{fig:MUSE-spectra-1}).
The source is a (likely interacting) pair with the source $\sim 1\farcs5$ to
the west, MUSE\#24, at the same redshift ($\Delta v \approx 76 $ km~s$^{-1}$).

ASPECS-LP.3mm.02: CO(2-1) at $z=1.317$.  Detected in both \OII\ and continuum
in MUSE.  This source is also detected in continuum at 1~mm and 5~cm
\citep[][UDF16]{Dunlop2017, Rujopakarn2016}.  \Oiia\ is severely affected by a
sky-line complicating the redshift and line-flux measurement.  We remeasure the
cataloged redshift for this source, which is used to compute the velocity
offset with CO(2-1) (\autoref{tab:CO-line-iden}).  Since we cannot confidently
recover the full \OII\ flux, we do not include this source in the analyses of
\autoref{sec:star-formation-rates} and \autoref{sec:metallicities-at-1.0}.

ASPECS-LP.3mm.03: CO(3-2) at $z=2.454$.  Photometric redshift indicates $z=2-3$
\citep{Skelton2014, Rafelski2015}, perfectly in agreement with the detection of
CO(3-2) at $z=2.45$.  The source is faint ($m_{\mathrm{F775W}} > 27$ mag) and
an extraction of the MUSE spectrum yields essentially no continuum signal (see
\autoref{fig:MUSE-spectra-1}).  This supports a redshift solution between
$z=2-3$, where no bright emission lines lie in the MUSE spectral range (see
\autoref{sec:muse}).  Beside there being little continuum in the spectrum,
there are no spectral features (in particular emission lines) indicative of a
lower redshift (\OII\ at $z=1.30$) or higher redshift (\Lya\ at $z= 3.60$)
solution.  Detected in continuum at 1~mm and 5~cm \citep[][UDF4]{Dunlop2017,
  Rujopakarn2016}.

ASPECS-LP.3mm.04: CO(2-1) at $z=1.414$. MUSE spectrum shows \OII\ and (weak)
continuum.  Detected in continuum at 1~mm and 5~cm \citep[][UDF6]{Dunlop2017,
  Rujopakarn2016}.

ASPECS-LP.3mm.05: CO(2-1) at $z=1.550$.  A massive ($M_{*} \approx 10^{11.5}$
\Msun) galaxy and an X-ray classified AGN \citep[][\#748]{Luo2017}.  It was
also detected by the ASPECS-Pilot in 1~mm continuum (C2, \citealt{Aravena2016};
cf. \citealt{Dunlop2017}), in CO(2-1) and also CO(5-4) and CO(6-5)
\citep[ID.3,][]{Decarli2016}, and in 5~cm continuum
\citep[][UDF8]{Rujopakarn2016}.  NIR spectroscopy from the SINS survey
\citep{ForsterSchreiber2009} reveals \Halpha, confirming the redshift we also
find from MUSE, based on the \FeII\ and \MgII\ absorption features.

ASPECS-LP.3mm.06: CO(2-1) at $z=1.095$.  Part of an overdensity in the HUDF at
the same redshift.  Rich star-forming spectrum in MUSE with a wealth of
continuum and emission features.  Detected in X-ray, but not classified as an
AGN \citep[][\#749]{Luo2017}.

ASPECS-LP.3mm.07: CO(3-2) at $z=2.696$.  Photometric redshift indicates $z=2-3$
\citep{Skelton2014, Rafelski2015}, perfectly in agreement with the detection of
CO(3-2) at $z=2.69$.  The source is faint ($m_{\mathrm{F775W}} > 27$ mag) and a
reextraction of the MUSE spectrum yields essentially no continuum signal (see
\autoref{fig:MUSE-spectra-2}).  This supports a redshift solution between
$z=2-3$, where no bright emission lines lie in the MUSE spectral range (see
\autoref{sec:muse}).  Beside there being little continuum in the spectrum,
there are no spectral features indicative of a lower redshift or higher
redshift solution (cf. ASPECS-LP.3mm.03).  There is reasonably close proximity
between ASPECS-LP.3mm.07 and 09 at $z=2.69$, which are separated by only
$\sim 7\farcs5$ (60 kpc at that redshift).  This object is one of the brightest
sources in the HUDF at 1~mm (UDF2; \citealt{Dunlop2017}) and also detected at
5~cm \citep{Rujopakarn2016}.

ASPECS-LP.3mm.08: CO(2-1) at $z=1.382$.  The source has a more complex
morphology which was already discussed in the ASPECS-Pilot program
(\citealt{Decarli2016}, see their Fig. 3).  The CO emission is spatially
consistent with a system of spiral galaxies.  MUSE reveals that the south-west
spiral is in the foreground at $z=1.087$.  Careful examination of the MUSE cube
reveals \OII\ emission matching the CO redshift in an arc north of the galaxies
and possibly towards the south-west, which is $\sim 1.8\arcsec$ away of peak of
the CO emission ($\sim 15$ kpc at the redshift of the source).  A potential
scenario is that the north-east spiral galaxies is the background source, in
which case the ionized gas emission of the spiral is completely obscured by the
disk of the (south-west) foreground spiral.  This is consistent with the
spatial position of the CO emission.  An alternative scenario is that of a
third disk galaxy harboring the CO reservoir, which is completely hidden from
sight by the spiral galaxies in the foreground, except for the structures seen
in the north and east.  We note that resolved SED fitting of this source was
recently performed by \cite{Sorba2018}, assuming the foreground redshift for
the entire system.  A clear break can be seen in the sSFR (their Figure 1.)
for the northern-arm and possibly also a south-west arm; consistent with
locations where \OII-emission is seen.  For the purpose of this paper, we
associate the north-east spiral with ASPECS-LP.3mm.08 and the south-west spiral
with ASPECS-LP-MP.3mm.02, but we note that this is uncertain in the case of
ASPECS-LP.3mm.08.  Given the limited flux we observe from the ionized gas, we
do not discuss this source in that context.

ASPECS-LP.3mm.09: CO(3-2) at $z=2.698$.  Photometric redshift indicates $z=2-3$
\citep{Skelton2014, Rafelski2015}, perfectly in agreement with the detection of
CO(3-2) at $z=2.69$.  The source is faint ($m_{\mathrm{F775W}} > 27$ mag), yet,
UV absorption features at $z=2.695$, matching the expected redshift of CO(3-2)
at $z=2.698$, are found in the MUSE spectrum at the position of the source (see
\autoref{fig:MUSE-spectra-3}).  The features arise in a source (MUSE\#6941)
$\sim 0\farcs8$ to the north ($\sim 6.5$ kpc at $z\sim2.7$).  The spectrum of
the northern source reveals a superposition of the $z=2.695$ source with a
foreground galaxy at $z = 1.555$.  This is also suggestive from the morphology
in \emph{HST}, which shows a redder central clump for the northern source.
Given the potential proximity of the two sources, both spatially and
spectrally, ASPECS-LP.3mm.09 could be part of a pair of galaxies with the
source to the north.  Notably, ASPECS-LP.3mm.09 is also detected as an X-ray
AGN; \citealt[][\#865]{Luo2017}.  Note there is also reasonably close proximity
between ASPECS-LP.3mm.07 and 09 at $z=2.69$, which are separated by only
$\sim 7\farcs5$ (60 kpc at that redshift).  One of the brightest sources in the
HUDF at 1~mm (UDF1; \citealt{Dunlop2017}), also detected at 5~cm
\citep{Rujopakarn2016}.

ASPECS-LP.3mm.10: CO(2-1) at $z=1.037$.  The lowest redshift detection.
Features a close star-forming companion at the same redshift.  The MUSE
spectrum shows continuum with both absorption and emission line features
(\OII).  We reextract the spectrum with a new segmentation map to recompute the
redshift and to minimize blending of the \OII\ flux from the close companion at
slightly different redshift.  The \OII\ line is detected in the source, but
given the residual deblending uncertainties we do not take into it into account
in the analyses of \autoref{sec:star-formation-rates} and
\autoref{sec:metallicities-at-1.0}.

ASPECS-LP.3mm.11: CO(2-1) at $z=1.096$.  Part of the overdensity in the HUDF at the
same redshift.  MUSE reveals a rich star-forming spectrum with stellar
continuum and both absorption and emission (\OII, \NeIII) features.

ASPECS-LP.3mm.12: CO(3-2) at $z=2.574$.  Detected in 1~mm continuum (C4;
\citep{Aravena2016}) and an X-ray AGN \citep[][\#680]{Luo2017}.  The source
contains a CO line at $96.76$ GHz.  The optical counterpart shows red colors in
\emph{HST} and features a blue component towards the north.  The source is
considered to be a single galaxy in most photometric catalogs
\citep[e.g.,][]{Skelton2014, Rafelski2015}.  However, the redshift from the
MUSE catalog for this source, $z=1.098$ (based on a confident \OII\ detection,
see \autoref{fig:MUSE-spectra-3}), is incompatible with being CO(2-1), which
would be at $z=1.383$.  Closer inspection of the source in the MUSE IFU data
reveals that the \OII\ emission is only originating from the blue clump to the
north of the source (see \autoref{fig:ASPECS12}).  A reanalysis of the MUSE
spectrum revealed weak absorption features that, when cross correlated with an
absorption line template, correspond a redshift $z=2.5738$.  Assuming that the
CO line is CO(3-2) instead, this independently matches the redshift from
ASPECS-LP exactly ($z=2.5738$).  To further confirm that the absorption
features are associated with ASPECS-LP.3mm.12, we spatially stacked
narrow-bands over all strong UV absorption features (without any preselection).
To construct the narrow-band, we sum the flux over each absorption feature
(assuming a fixed 7\AA\ line-width) and subtract the continuum measured in two
side bands offset by $\pm 10$\AA\ (same width in total).  We then stacked the
individual narrow-bands by summing the flux in each spatial pixel (note, the
same result is found when taking the mean or median).  The stacked absorption
features have S/N~$> 4$ and are co-spatial with the background galaxies and the
CO, confirming the detection of CO(3-2) at $z=2.5738$ (see
\autoref{fig:ASPECS12}).

ASPECS-LP.3mm.13: CO(4-3) at $z=3.601$.  Highest redshift CO detection.  It is
an F435W dropout and the photometric redshifts for this source consistently
suggest that it lies in the $z=3-4$ range \citep{Skelton2014, Straatman2016},
with $z_{\rm BPZ}=3.67_{-0.24}^{+0.74} $ \citep{Rafelski2015}.  These all
suggest a detection of CO(4-3) at $z=3.601$.  In order the spectroscopically
confirm this redshift, we extract a MUSE spectrum at the position of the
source.  The strongest UV emission line observed by MUSE at these redshifts is
\Lya, while it also covers the much weaker \CIII\ line.  Both are not detected
in the spectrum of ASPECS-LP.3mm.13.  The non-detection of \CIII\ at the 10~h
depth of the mosaic is understandable, as robustly detecting \CIII\ at these
redshifts is challenging (see \citealt{Maseda2017} for a in-depth discussion,
which finds the highest redshift detection of \CIII\ in the deep 30~h MUSE data
to be at $z\sim2.9$).  Unfortunately, at $z=3.601$ the expected position of
\Lya\ in MUSE falls close to the \Oisky\ skyline
(\autoref{fig:MUSE-spectra-4}), which could explain why it is not detected.
Furthermore, the source is likely to have a significant dust content in which
case no \Lya\ emission may be expected at all.  Nevertheless, while at
$m_{\mathrm{F775W}} = 26.4$, the spectrum does not reveal emission or
absorption lines compatible with a solution for CO(2-1) at $z=1.30$ or CO(3-2)
at $z=2.45$, which suggests a higher redshift solution is appropriate for
ASPECS-LP.3mm.13 (in agreement with the photo-z).  In summary, the combined
evidence of the photometric redshifts indicating $z\sim3.5$ and the lack of a
lower redshift solution from MUSE makes the case for the detection of CO(4-3)
at $z=3.601$ in ASPECS-LP.3mm.13.

ASPECS-LP.3mm.14: CO(2-1) at $z=1.098$.  Part of an overdensity in the HUDF.  MUSE
reveals a rich spectrum with continuum, absorption and a range of emission
lines (among which \OII, \NeIII\ and Balmer lines).

ASPECS-LP.3mm.15: CO(2-1) at $z=1.096$.  Part of an overdensity in the HUDF at
the same redshift.  The source lies in a very crowded part of the sky with
multiple galaxies at different redshifts overlapping in projection.  Detected
in X-rays, classified as AGN \citep[][\#689]{Luo2017}.  Source was also covered
by the ASPECS-Pilot program and detected in CO(2-1) and CO(4-3)
\citep[][ID.5]{Decarli2016}.

ASPECS-LP.3mm.16: CO(2-1) at $z=1.294$.  Shows a disk-like morphology.  MUSE spectrum
reveals a stellar continuum with absorption, as well as emission lines (\OII\
and \NeIII).

ASPECS-LP-MP.3mm.01: CO(2-1) at $z=1.096$.  Part of an overdensity in the HUDF
at the same redshift.  MUSE spectrum shows stellar continuum with absorption,
as well as emission lines (\OII).

ASPECS-LP-MP.3mm.02: CO(2-1) at $z=1.087$.  Foreground galaxy to
ASPECS-LP.3mm.08, also described in \cite{Decarli2016}.  See ASPECS-LP.3mm.08
for a further description.

\section{\textsc{Magphys fits for all CO detected galaxies}}
\label{sec:magphys-fits}
We performing SED fitting with \textsc{magphys} for all ASPECS-LP galaxies, as
described in detail in \autoref{sec:multi-wavel-data}.  The following bands are
considered in the SED fitting of the ASPECS-LP galaxies: U$_{38}$
(0.37\micron), IA$_{427}$ (0.43\micron), F435W (0.43\micron), B (0.46\micron),
IA$_{505}$ (0.51\micron), IA$_{527}$ (0.53\micron), V (0.54\micron), IA$_{574}$
(0.58\micron), F606W (0.60\micron), IA$_{624}$ (0.62\micron), IA$_{679}$
(0.68\micron), IA$_{738}$ (0.74\micron), IA$_{767}$ (0.77\micron), F775W
(0.77\micron), I (0.91\micron), F850LP (0.90\micron), J (1.24\micron), tJ
(1.25\micron), F160W (1.54\micron), H (1.65\micron), tK$_{\rm s}$
(2.15\micron), K (2.21\micron), IRAC (3.6\micron, 4.5\micron, 5.8\micron,
8.0\micron), MIPS (24\micron), PACS (100\micron\ and 160\micron) and ALMA Band
6 (1.2~mm) and Band 3 (3.0~mm).

\begin{figure*}[t]
\includegraphics[width=\textwidth]{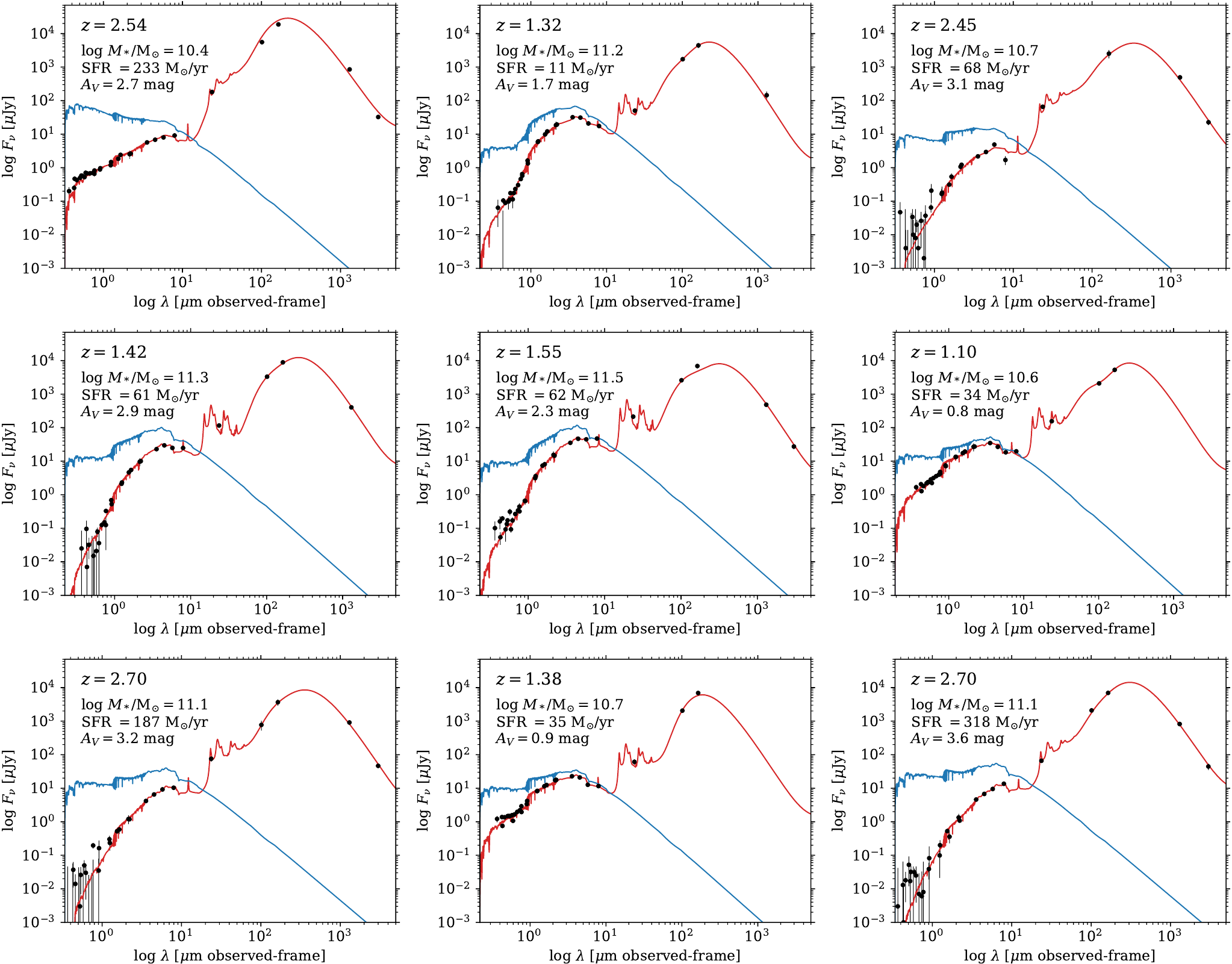}
\caption{Spectral energy distributions (SEDs) for all the ASPECS-LP CO detected
  sources from the line search (first sixteen) and MUSE redshift prior based
  search (last two).  The black points are the observed photometry.  The
  overall best fit SED from \textsc{magphys} is shown by the red line, while
  the the model of the unattenuated stellar emission is shown in blue.  The
  redshift and median values of the posterior likelihood distribution of the
  stellar mass ($M_{*}$), star formation rate (SFR) and visual attenuation
  ($A_{V}$) are indicated in each panel. \label{fig:fig_sed}}
\end{figure*}

\begin{figure*}[t]
  \includegraphics[width=\textwidth]{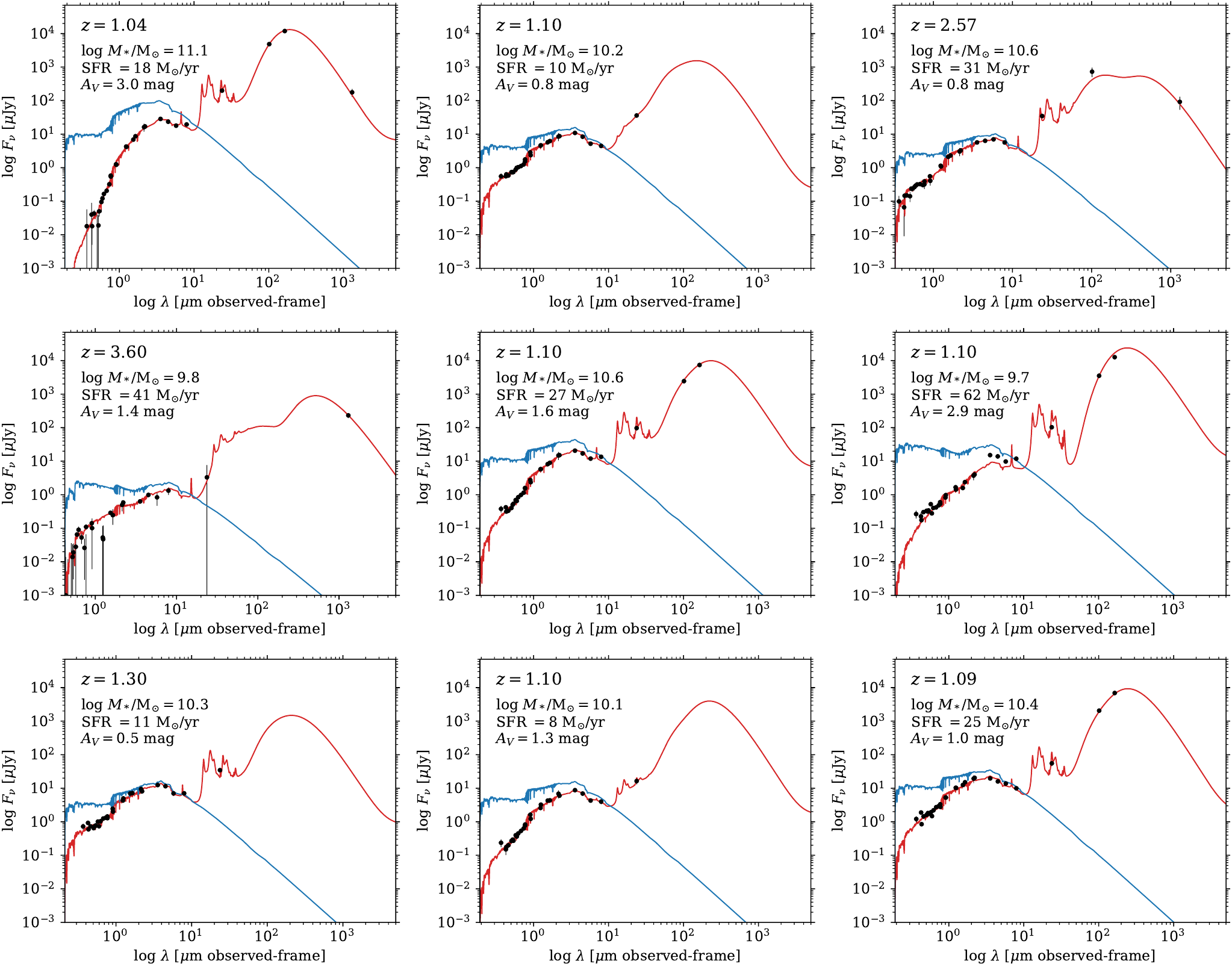}
\caption{Continuation of \autoref{fig:fig_sed}. \label{fig:fig_sed2}}
\end{figure*}

\bibliographystyle{aasjournal}
\bibliography{library.bib}

\end{document}